\documentclass[preprint,showpacs,preprintnumbers,amsmath,amssymb]{revtex4}
\usepackage{graphicx}
\usepackage{color}
\usepackage{bm}

\begin{document}

\preprint{APS/123-QED}

\title{Theory and simulation of the confined Lebwohl-Lasher model}

\author{R. G. Marguta}
\affiliation{Instituto de Qu\'{\i}mica-F\'{\i}sica Rocasolano, CSIC, Serrano 119, E-28006, Madrid, Spain}

\author{Y. Mart\'{\i}nez-Rat\'on}
\affiliation{Grupo Interdisciplinar de Sistemas Complejos (GISC),
Departamento de Matem\'{a}ticas,Escuela Polit\'{e}cnica Superior,
Universidad Carlos III de Madrid, Avenida de la Universidad 30, E--28911, Legan\'{e}s, Madrid, Spain}

\author{N. G. Almarza}
\affiliation{Instituto de Qu\'{\i}mica-F\'{\i}sica Rocasolano, CSIC, Serrano 119, E-28006, Madrid, Spain}

\author{E. Velasco}
\affiliation{Departamento de F\'{\i}sica Te\'orica de la Materia Condensada
and Instituto de Ciencia de Materiales Nicol\'as Cabrera,
Universidad Aut\'onoma de Madrid, E-28049 Madrid, Spain}

\date{\today}

\begin{abstract}
We discuss the Lebwohl-Lasher model of nematic liquid crystals in a confined
geometry, using Monte Carlo simulation and mean-field theory. 
A film of material is sandwiched between two planar, parallel 
plates that couple to the adjacent spins via a surface strength
$\epsilon_s$. We consider the cases where the favoured alignments at
the two walls are the same (symmetric cell) or different 
({asymmetric cell}). In 
the latter case, we demonstrate the
existence of a {\it single} phase transition in the slab for all 
values of the cell thickness. This transition has been observed before 
in the regime of narrow cells, where the two structures involved 
correspond to different arrangements of the nematic director. By studying
wider cells, we show that the transition is in fact
the usual isotropic-to-nematic (capillary) transition under confinement in the case of
antagonistic surface forces. We show results for
a wide range of values of film thickness, and discuss the phenomenology using a
mean-field model.
\end{abstract}

\pacs{61.30.Cz, 61.30.Hn, 61.20.Gy}

\maketitle

\section{Introduction}
\label{introduction}

The Lebwohl-Lasher lattice spin model \cite{LL}
is an important model to understand the
formation of the nematic phase in mesogenic materials. It provides
qualitatively correct predictions and, in some cases, 
even quantitative information about nematic properties \cite{LL1,Zhang}. 
There is renewed interest in the model as regards the behaviour of nematic films 
and the nature of the orientational phase transition. Also recently
the confined model has been analysed in hybrid geometry \cite{Zann1,Zann2}.

\begin{figure}
\includegraphics[width=3.0in]{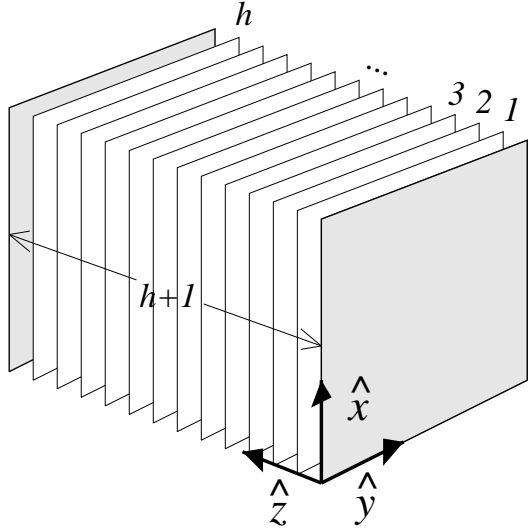}
\caption{Schematic representation of the confined Lebwohl-Lasher model. $h$ is the number
of spin layers sandwiched between the two external plates (shaded). 
$h+1$ is the film thickness in units of the cubic lattice parameter. 
The units vectors along the Cartesian coordinates are indicated.}
\label{model}
\end{figure}

In the model, spin unit vectors $\hat{\bm s}$ 
are located at the sites of a cubic lattice of lattice parameter $a$.
Nearest-neighbour (NN) spins interact via a potential energy 
$-\epsilon P_2(\cos{\gamma})$,
where $\epsilon$ is a coupling parameter ($\epsilon > 0$), and
$\cos{\gamma}=\hat{\bm s}\cdot\hat{\bm s}^{\prime}$, with $\gamma$ the 
relative angle between the two spins. $P_2(x)$ is the second-degree
Legendre polynomial. 
In the confined model (see Fig. \ref{model}), parallel spin layers, 
$h$ in number, 
are sandwiched between two planar, parallel
plates (slit pore geometry), each formed by frozen spins that
interact with spins in the first and last layers (those adjacent to
the plates) also with the same potential, but with a (surface) coupling constant $\epsilon_s$.
The Hamiltonian of the model is then
\begin{eqnarray}
{\cal H}=-\epsilon\sum_{\hbox{NN}} P_2(\hat{\bm s}\cdot\hat{\bm s}^{\prime})
-\epsilon_s^{(1)}\sum_{\hbox{\small first layer}} P_2(\hat{\bm m}_1\cdot\hat{\bm s})
-\epsilon_s^{(2)}\sum_{\hbox{\small last layer}} P_2(\hat{\bm m}_2\cdot\hat{\bm s})
\end{eqnarray}
where the first sum extends over all distinct NN spins,
and the second and third only involve the spins in the first and last layers,
respectively.
{The surface coupling constants $\epsilon_s^{(i)}$, $i=1,2$, may or may not
be different for both plates. In the simulations to be presented
below, we take $\epsilon_s^{(1)}$ and $\epsilon_s^{(2)}$ to be identical
(in Section \ref{mean} the case of different constants will be considered)} 
but, in general, each plate is assumed to favour a different spin orientation (easy axis), 
$\hat{\bm m}_1$ or $\hat{\bm m}_2$.
The case $\hat{\bm m}_1=\hat{\bm m}_2$ is a particular
case, the symmetric cell, while $\hat{\bm m}_1\ne\hat{\bm m}_2$ is the asymmetric case,
{also called hybrid or twisted cell, depending on the actual orientation of the
axes}. Since the number of fluctuating spin layers is $h$, the cell width is $h+1$ in units of
the cubic lattice parameter $a$. The symmetry of the confined model implies that its
properties only depend on the scalar $\hat{\bm m}_1\cdot\hat{\bm m}_2$, and not
on the individual components of the easy axes. {In this respect, our cell is both
hybrid (a name reserved for the case where one of the axes is normal to its
surface, while the other is parallel) and twisted (a situation where the two axes
lie on the surface planes).}

The situation where $\hat{\bm m}_1\cdot\hat{\bm m}_2=0$ is very interesting, as the 
film will be subject to antagonistic but equivalent forces
at the plates, which create frustration. The nematic director can 
satisfy both surface forces by rotating across the slab, creating
an approximately linearly dependent, smoothly rotated director configuration (L phase),
which involves an elastic energy. There have been two recent Monte Carlo 
(MC) simulations of this model \cite{Zann1,Zann2}, motivated by previous works that 
indicated the existence of a step-like slab configuration 
(S phase, sometimes called biaxial or exchange-eigenvector phase) in which the 
director is constant except in a thin central
region, where it rotates abruptly between the two favoured orientations
\cite{5,6,9,10}. These preliminary works, {along with a more recent one on the
twisted cell but with $0<\hat{\bm m}_1\cdot\hat{\bm m}_2\le 1$} \cite{Bisi}, are
based on Ginzburg-Landau-type models, and
predict a L to S (LS) phase transition that was confirmed by the MC studies.
{A recent analysis of a hybrid cell using a surface-force apparatus may have
detected this transition experimentally \cite{Zappone}.}
But the nature of the transition, the effect of plate separation, and
especially the relationship between the LS transition and the bulk behaviour 
(i.e. isotropic-nematic, or IN transition) have not been addressed in MC 
simulations. Some work on related, continuum nematic-fluid slabs under hybrid
conditions, analysed by means of density-functional theory, have appeared 
recently and partially answered some of these questions \cite{us,Paulo}.

The MC results of Ref. \cite{Zann1} only presented a partial scenario of the
problem. As mentioned, the connection of the LS transition with the bulk isotropic-nematic
phase transition remained obscure, and the effect of plate strength
$\epsilon_s$ and the regime of very small separation were not
explored. In a recent paper \cite{Zann2}, the authors add some confusion 
to the problem by implicitely stating that there is a {\it second} transition 
in the slab, of unknown origin, inferred from a weak signal in the specific 
heat of the slab, as obtained from their MC simulations.

In the present paper we perform careful MC simulations on
the hybrid cell with {$\epsilon_s^{(i)}=\epsilon$, $i=1,2$. These simulations
will be supplemented by mean-field (MF) theoretical results, where cases with
$\epsilon_s^{(1)}\ne\epsilon_s^{(2)}$ will also be considered.}
We obtain the LS phase transition from
specific-heat data obtained from long MC simulation runs, and extend 
the analysis to very small separations, including the case of 
a single spin layer. No additional transitions are observed in
our simulations. The connection with the bulk IN
transition is established by performing simulations on thicker nematic films, 
supplemented by MF calculations. The available evidence indicates
that there is a single transition line in the phase
diagram, namely the LS transition, and that this transition coincides with
the capillary IN transition in the confined system, which is connected
with the bulk IN transition as the plate separation $h\to\infty$.

In the remaning sections we first discuss the MC simulation techniques
(Section \ref{simul}), and then show the results obtained 
for the case of symmetric (Section \ref{sym}) and asymmetric (Section \ref{asym}) plates.
The MF model and its results are shown in Section \ref{mean}, which includes a discussion
on the macroscopic approach (Kelvin equation) for this problem. 
{The connection with the wetting properties is also discussed.
A short discussion on the general picture and on relation of the present results with those of Ref. 
\cite{Zann1} is given in Section \ref{discussion}.}
Conclusions are presented in Section \ref{conclusions}. Some details of the macroscopic
model can be found in the Appendices.

\section{Monte Carlo technique}
\label{simul}

Let us take each of the $h$ layers to consist of $L\times L$ spins.
The total number of spins is then $N=hL^2$.
The MC simulation runs include two types of moves: one-particle
orientational moves, and cluster moves.
The one-particle orientational moves are carried out using the
standard algorithm for linear molecules described in Ref. \cite{Frenkel_book}.
The cluster moves are performed by means of the usual bonding criteria for
NN particles \cite{Kunz,Priezjev}.
The presence of the wall-particle interactions imposes
some restrictions on the possible reflections that can be used to carry
out the cluster moves.
Notice, however, that the total energy is invariant with
respect to a simultaneous change of sign of all the $x$ components of the
particle orientations.  The same property applies to the $y$ and $z$
components. Therefore, in our realization
of the cluster algorithm, we choose at random the component
($s_x$, $s_y$ or $s_z$) that will eventually flip. Then, we test the creation of bonds between
every NN pair of particles
by taking into account the change of interaction energy if only the
coordinate of one of the
particles of the pair is flipped, the bonding probability
being \cite{Kunz,Priezjev,Wolff,Swendsen}:
\begin{equation}
b_{ij} = 1 - \exp \left\{
\min \left[ 0,
\frac{- 6 \epsilon}{kT} \left( s_{\alpha i} s_{\alpha j}
 \hat{\bm s}_i \cdot \hat{\bm s}_j - s_{\alpha i}^2  s_{\alpha j}^2
\right)
\right]
\right\};
\label{bonding}
\end{equation}
where $\alpha=\{x,y,z\}$ is the chosen direction for the reflections, $k$ is Boltzmann's
constant and $T$ is the temperature.
Once all the possible bonds have been tested, the actual bond realization is used
to distribute the system in several clusters of particles.
The cluster move is then performed following the Swendsen-Wang strategy \cite{Swendsen}:
each cluster is flipped (or not flipped) with probability one half.

The simulations were organized in blocks, each block containing 15000 cycles.
%{\color{red} A cycle consists of trial one-particle orientational moves and one-cluster 
{A cycle consists of trial one-particle orientational moves and one-cluster 
move}. After an equilibration period
of about 150 blocks, we calculate averages 
%{\color{red}over 175 additional blocks} of the
{over 175 additional blocks} of the
potential energy per particle $u$, and the eigenvectors and eigenvalues
of different realisations of the local 
Saupe tensor ${\cal Q}_i$, at each
plane $i=1,...,h$. This tensor has components
\begin{equation}
{\left({\cal Q}_i\right)_{\alpha\beta}=
\frac{1}{L^2}\sum_{k\in{i\hbox{\tiny th layer}}}\frac{1}{2}\left(3s_{\alpha k}s_{\beta k}-
\delta_{\alpha\beta}\right),\hspace{0.6cm}\alpha,\beta=x,y,z},
\end{equation}
%{\color{red}where the sum extends over all spins of the $i$th plane, $N_i$ in number.
{where the sum extends over all spins of the $i$th plane, $L^2$ in number.
%$\left<...\right>$ denotes a thermal average over spin configurations.
The local tensor ${\cal Q}_i$, defined in each layer, is diagonalised, providing 
eigenvalues $P_i$, $-(P_i-B_i)/2$ and $-(P_i+B_i)/2$. The first, associated with the 
$x$ direction in the proper frame (i.e. the frame where ${\cal Q}_i$ is diagonal),
which coincides with the local nematic director $\hat{\bm n}_i$, is the local uniaxial nematic
order parameter, whereas $B_i$ is the biaxial nematic order parameter.
The orientation of the proper frame with respect to the lab (plate-fixed) frame at each plane 
is given by the tilt angle $\phi_i$, which describes the
director orientation in the $xy$ plane (spanned by the plate orienting fields) and coincides
with the angle between the $x$ axes of the two frames. We define $-\pi<\phi_i<\pi$. For the symmetric cell, 
we take $\hat{\bm m}_1=\hat{\bm m}_2=\hat{\bm x}$, and $\left<\phi_i\right> \simeq 0$.}
$\left<...\right>$ denotes a thermal average over spin configurations.
For the asymmetric cell, with $\hat{\bm m}_1\cdot\hat{\bm m}_2=0$
(we take $\hat{\bm m}_1=\hat{\bm x}$ and $\hat{\bm m}_2=\hat{\bm y}$), 
we compute, for each layer $i$, the angle $\phi_i$ as:
\begin{equation}
\cos{\phi_i}= \left ( \frac{\langle n_{xi}^2 \rangle}
{\langle n_{xi}^2 \rangle + \langle n_{yi}^2 \rangle} \right)^{1/2},
\end{equation}
where $n_{xi}$ and $n_{yi}$ are the $x$ and $y$ components of $\hat{\bm n}_i$ {
(thermal averages of local quantities at sites lying in the same plane are identical by symmetry).}
{Note that, due to the high symmetry of the spin interaction, only one 
deformation mode of the angle $\phi_i$ is possible in the cell (so that splay, bend and twist are
equivalent; see Appendix \ref{elastic})}. To analyze possible second-order phase transitions,
we also compute an additional order parameter, $P_{xy}$, with
\begin{equation}
P_{xy} = \left< \frac{1}{N} \left| \sum_{k=1}^{N} s_{xk}s_{yk} \right|\right>,
\end{equation}
and, for each plane $i$, the local order parameters:
\begin{equation}
(P_{xy})_i = 
 \left< \frac{1}{L^2} \left| \sum_{k\in i\hbox{\tiny th layer}} s_{xk}s_{yk} \right|\right>,
\end{equation}
The order parameter $P_{xy}$ describes the global orientation of the
particles in the plane of the interacting fields and is related
to {the thermal average of} the absolute value of one of the off-diagonal 
elements of the Saupe tensor by
$P_{xy}=(2/3) \left<\left|{\cal Q}_{xy}\right|\right>$. 
Likewise, global uniaxial and biaxial order parameters $P$ and $B$ can be defined:
\begin{equation}
P = \frac{1}{h}\sum_{i=1}^hP_i,\hspace{0.4cm} B = \frac{1}{h}\sum_{i=1}^hB_i.
\end{equation}
Notice that these global order parameters do not correspond to those that could
be computed by diagonalising the global Saupe tensor.
The following relation holds between $P_i$, $B_i$, $(P_{xy})_i$ and $\phi_i$ locally (at each plane):
\begin{equation}
\left(P_{xy}\right)_i=\frac{1}{2}\left(P_i-\frac{B_i}{3}\right)\left|\sin{2\phi_i}\right|.
\end{equation}
Therefore, $\left(P_{xy}\right)_i$ reflects the variations of both the nematic order parameters 
$P_i$ and $B_i$, and of the director tilt angle $\phi_i$, across the slab. The LS transition 
can be monitored in principle by the changes with temperature of the global
order parameters $P$, $B$ and $P_{xy}$. As we will see, our simulations indicate that the transition
in the confined slab has a continuous nature in the range of pore widths explored, so that 
these order parameters do not undergo discontinuities, but are singular in their derivatives.
The associated singularities are washed out in our (necessarily) finite-size simulations. 
In fact, the finite-size dependence of the order parameters is very weak, and simulations
on systems with large lateral sizes, along with a proper finite-size scaling analysis, are
required. However, relevant response functions provide a more clear-cut signature of the
transition. We have focused on the excess heat capacity per spin,
$c_{v}=(\partial u/\partial T)_h$, with $u=\left<{\cal H}\right>/N$ the average internal energy per
spin. In the simulations $c_v$ is obtained from the fluctuations in the energy. The phase-transition 
temperature will be located as that temperature at which $c_{v}$ reaches a maximum value.

In order to locate the maximum in the heat capacity we use the
synthetic method proposed by de Miguel \cite{demiguel08},
which we briefly describe in the following.
Let us consider that $c_{v_{i}}^{(0)}$ are the output values of
the heat capacity and $\Delta c_{i}$ their
associated statistical errors as obtained from MC simulations
at input temperatures $T_{i}$, $i=1, \ldots n$.
Usually we fit $c_{v_{i}}^{(0)}$ to
a polynomial of order $M$ in $T$, $c_{v}(T)= \sum_{i=1}^M a_{i}T^{i-1}$.
We search the maximum of this
polynomial function by computing the value of the temperature, $T_m$,
for which the derivative of the heat capacity with respect to
the temperature is zero,
then we calculate $c_{v_{m}}=c_{v} (T_{m})$.
The synthetic method consists of the following steps:
\begin{itemize}
\item Generate synthetic sets of $n$ data points,
$c_{v_{i}}^{(k)} = c_{v_{i}}^{(0)} + \xi_{i}$, where
$\xi$ is a random number drawn from a Gaussian
distribution with zero mean value and
standard deviation $\Delta c_i$.
\item Find the fitting coefficients $a_{i}^{(k)}$ and
calculate $c_{v_{m}}^{(k)}$ corresponding to each synthetic set.
The set of maximum heat capacities follows a
Gaussian distribution, and we determine the mean value
$c_{v}^{\rm max}(L,h)$.
\end{itemize}
Note that for each synthetic set generated we calculate
$T_{m}^{(k)}$. This set of temperatures will also follow a
Gaussian distribution, so we can determine the mean value, which will
be denoted by $T_c(L,h)$.

\section{Results for the symmetric cell}
\label{sym}

First we report on the case of symmetric plates, $\hat{\bm m}_1=
\hat{\bm m}_2$. This case has been investigated in detail by various 
authors, using MC simulation \cite{Doug,Todos}, MF 
theory \cite{Todos1,Todos2} and renormalisation-group (RG) techniques \cite{Tim}.
The cases $\epsilon_s>0$, favouring positive 
order parameter, and $\epsilon_s<0$, favouring negative order
parameter, were considered. Here we focus on the
first, using $\epsilon_s^{(1)}=\epsilon_s^{(2)}=\epsilon$. 
Mean-field models predict a weak 
first-order transition, and a terminal plate separation $h_t$
below which the capillary isotropic-nematic transition disappears.
The plain MF model gives $h_t=14$, whereas a Bethe model,
including two-spin correlations \cite{Todos2}, increases the value up to $h_t=21$.
Assuming a monotonic variation due to higher-order fluctuations,
we may expect $h_t\agt 21$. Simulations and RG calculations
predict a continuous transition, in disagreement with MF results.

\begin{figure}
\includegraphics[width=3.6in]{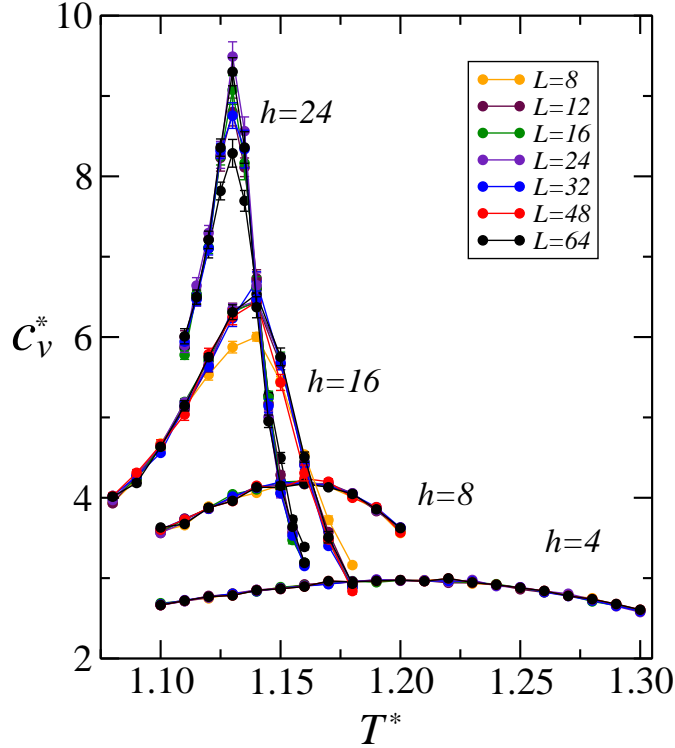}
\caption{\small{(Colour online). Excess heat capacity per spin in reduced units, $c_v^*$, as a function
of reduced temperature $T^*$, for the symmetric case and for various plate separations 
(indicated as labels). Several lateral sizes, given in the inbox, are considered in each case.}}
\label{Cvsym}
\end{figure}

Our own simulation results were based on long runs using the special
techniques described in the previous section. Our results, obtained for 
plate separations $h\le 24$, are not compatible with the existence of a
phase transition. Fig. \ref{Cvsym} shows the behaviour of the
heat capacity per spin, $c_v^*=c_v/k$, as a function
of reduced temperature $T^*=kT/\epsilon$. 
Various plate separations $h$ are shown. In each case an analysis of how the lateral size
of the sample $L$ affects the results has been done. We can
see that $c_v$ does not show any significant dependence with $L$
(provided that $L>h$) as $L\to\infty$,
even for $h=24$. Therefore, we may expect $h_t>24$.\\

\section{Results for the hybrid cell}
\label{asym}

The hybrid cell is the main focus of our work. For this cell we chose ${\bm m}_1=\hat{\bm x}$ and
${\bm m}_2=\hat{\bm y}$. We have simulated systems with different number of slabs for plate strength
$\epsilon_s^{(1)}=\epsilon_s^{(2)}=\epsilon$. 
In the following, detailed results are presented for
the cases $h=8$, which is representative of the LS 
phase transition within a narrow pore, and $h=1$, which is a special case. At the
end of the section the global phase diagram, spanning a wide range of values of $h$, will be
discussed. In particular, we show results for the case $h=32$, which illustrate the nature of the
LS phase transition in the regime of wide cells and are used to pinpoint the
main differences with respect to the regime of narrow cells.

\subsection{${\bf h=8}$}
\label{h8}

\begin{figure}
\centering
\includegraphics[height=16.0cm,angle=0 ]{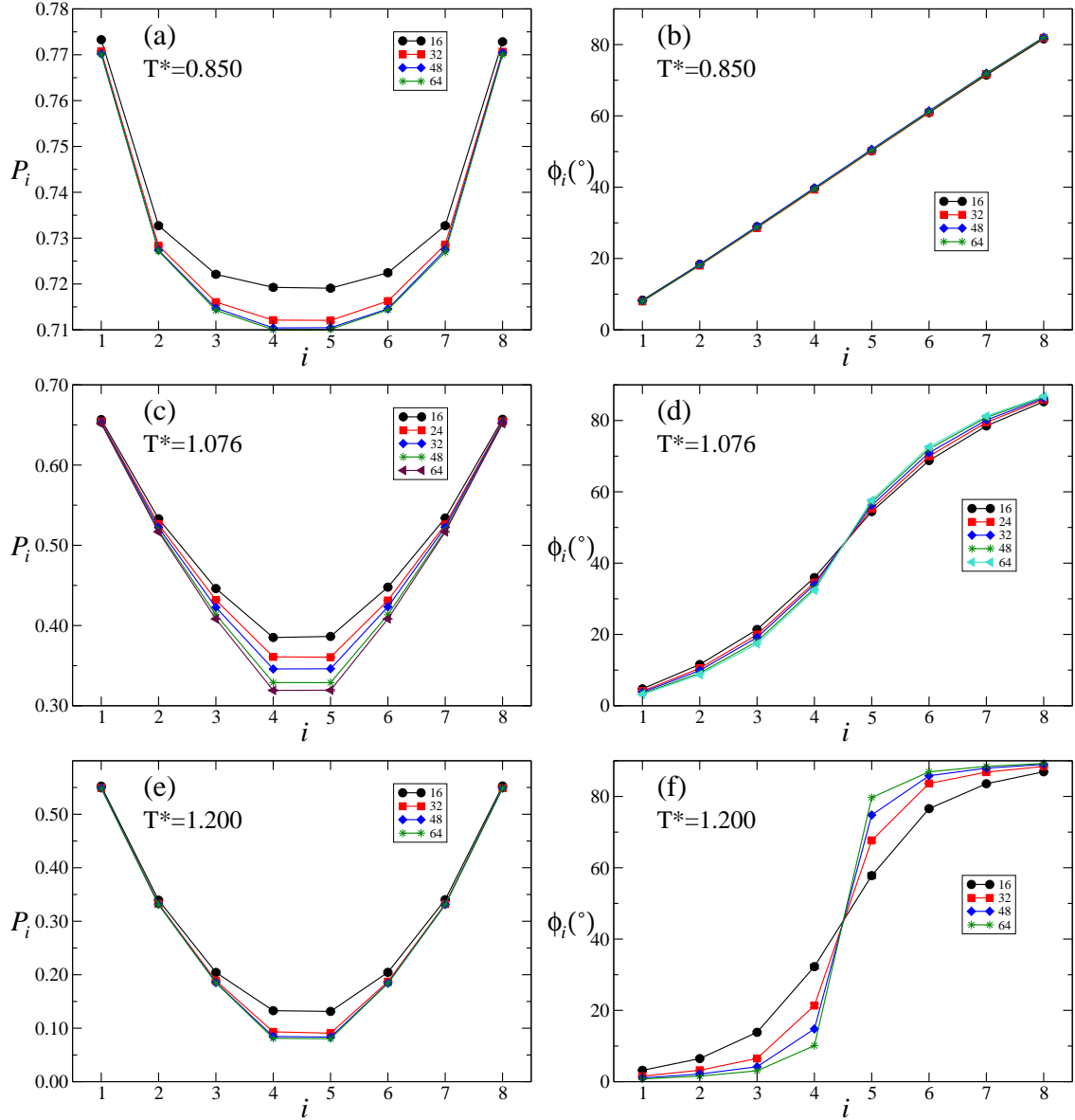}
\caption {\small{(Colour online). Nematic uniaxial order parameter $P_i$
(left panels) and tilt angle of the nematic director $\phi$ along the $z$ direction
for the slit pore with $h=8$. (a) and (b) $T^*=0.850$; (c) and (d) $T^*=1.076$; (e) and (f) $T^*=1.200$.
The lateral size $L$ used in the simulations is indicated in the inbox.}}
\label{fig.p}
\end{figure}

The uniaxial nematic order parameter {$P_i$ and the tilt angle $\phi_i$ profiles}
are plotted in Fig. \ref{fig.p} for different values of reduced temperature.
In agreement with earlier predictions found in the literature
\cite{Zann1,us,Paulo,Zann2},
the orientational structure changes {{\it continuously} or
{\it discontinuously} across the slab, depending on the temperature (obviously, one cannot
strictly talk about continuous or discontinuous functions in a discrete system; 
these are fuzzy adjectives that we ascribe to an interpolating function, passing through all
points in the profiles, that could reasonably be drawn in each case)}. For example,
the tilt angle clearly shows that, for the highest temperature, there is
a discontinuity in the centre of the slab, this change becoming steeper as
the system size is increased. This is the step-like (S) phase. By contrast, at 
low temperature, the orientation of the director changes smoothly from
$\hat{\bm x}$ to $\hat{\bm y}$: this is the linear-like (L) phase.
At higher or lower temperatures no additional structural changes are visible
in the order parameters or tilt angle. We conclude that there must be a
temperature $T_c$ at which the structure changes from the S to the L configuration
as a thermodynamic phase transition, and that, in view of the smooth variation 
of the profiles with temperature, one can assume this transition to be continuous.
Later we will provide evidence that, in the thermodynamic limit $L\to\infty$, 
the tilt-angle profile at the transition, corresponding to the situation depicted in
Fig. \ref{fig.p}(d), is actually a step function.

\begin{figure}
\centering
\includegraphics[height=12cm,angle=0 ]{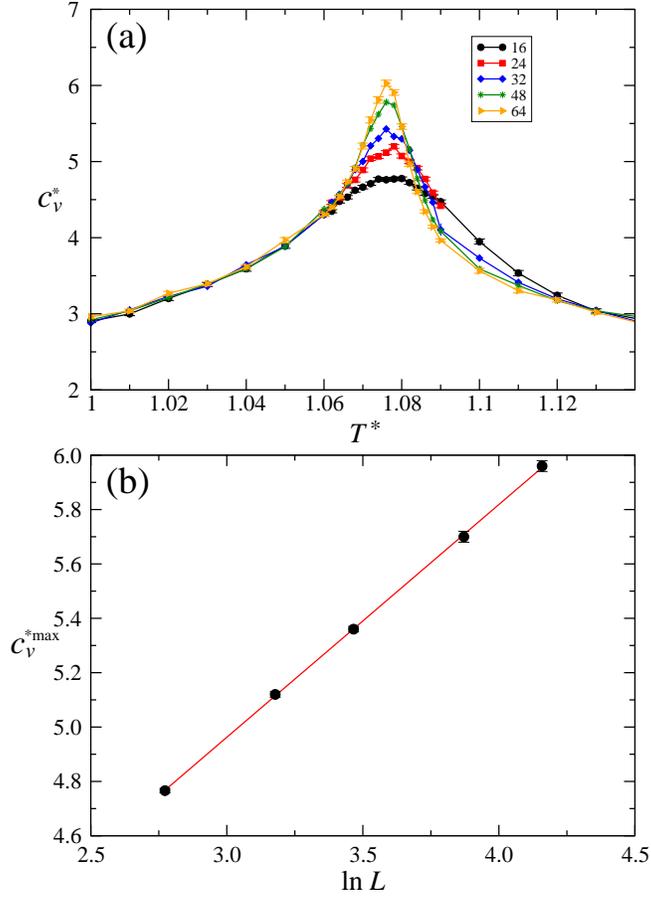}
\caption {\small{(Colour online). (a) Excess heat capacity per spin in reduced units, $c_v^*$, for the 
system with $h=8$, as a function of reduced temperature $T^*$ and for different lateral 
system sizes $L$ (indicated in the inbox).
(b) Maximum of the heat capacity per spin in reduced units, 
$c_v^{*\hbox{\tiny max}}$, for the system with $h=8$,
as a function of lateral system sizes $L$. The straight line is a linear fit.}}
\label{fig4}
\end{figure}

More information about the structural LS transition can be found by looking at the heat capacity.
The phase transition is signalled by a diverging maximum of the
heat capacity as the system lateral size is increased, Fig. \ref{fig4}(a).
The maximum exhibits a linear dependence with $\log{L}$, as shown in
Fig. \ref{fig4}(b). This dependence suggests that the confined Lebwohl-Lasher system 
under hybrid conditions {for the case $h=8$} presents a continuous transition belonging to the 
universality class of the two-dimensional Ising model \cite{Landau_book}.

\begin{figure}
\centering
\includegraphics[height=8cm,angle=0 ]{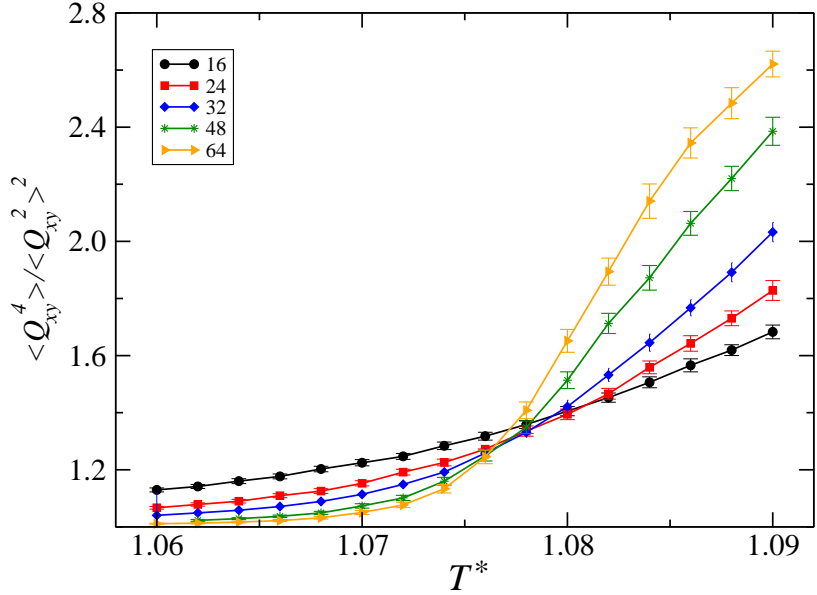}
\caption {\small{(Colour online). Dependence of the normalised fourth-order
cumulant $G_4=\left<Q_{xy}^4\right>/\left<Q_{xy}^2\right>^2$ with reduced temperature 
$T^*=kT/\epsilon$ and
lateral size $L$ for the system with $h=8$. The lateral size of the systems is quoted in the
legend.}}
\label{fig.cum}
\end{figure}

Such a hypothesis is fully supported by considering a cumulant analysis of spin
correlations in the $xy$ plane. Specifically, we define a global Saupe tensor as 
\begin{eqnarray}
{\cal Q}=\frac{1}{h}\sum_{i=1}^h{\cal Q}_i,
\end{eqnarray}
and focus on the tensor element ${\cal Q}_{xy}$. 
The finite-size dependence \cite{Landau_book} of the quantity
$G_4 \equiv \langle {\cal Q}_{xy}^4 \rangle / \langle  {\cal Q}_{xy}^2\rangle ^2$
turns out to be fully consistent with the proposed critical behavior.
%Such a hypothesis is fully supported by considering $P_{xy}$
%as an order parameter to analyze the transition,
%with a role similar to that of the magnetization in the Ising model.
%The finite-size dependence \cite{Landau_book} of the quantity
%$G_4 \equiv \langle {\cal Q}_{xy}^4 \rangle / \langle  {\cal Q}_{xy}^2\rangle ^2$
%is fully consistent with the proposed critical behavior.
The results for $h=8$ and different values of $L$ are presented in Fig. \ref{fig.cum}.
As expected, the different curves intersect at values of $G_4$ not too
far from the universal value $G_{4c} \simeq 1.168$ of the two-dimensional Ising
universality class for systems with $L_x=L_y$ and periodic boundary conditions
\cite{Salas_2000}. Then we assume the scaling relation \cite{Landau_book}
\begin{equation}
T_c(L,h) = T_c(h) + a L^{-1/\nu},
\label{eq.tclh}
\end{equation}
where the critical exponent has the value $\nu=1$  for
the two-dimensional Ising universality class \cite{Salas_2000},
and obtain the critical temperature of the transition 
as $T_c(h) = \lim_{L\rightarrow \infty} T_c(L,h)$.
For the particular value of cell thickness $h=8$ 
we obtain $T_c^*(h=8)=1.076(1)$.

\begin{figure}
\centering
\includegraphics[height=14.0cm,angle=0 ]{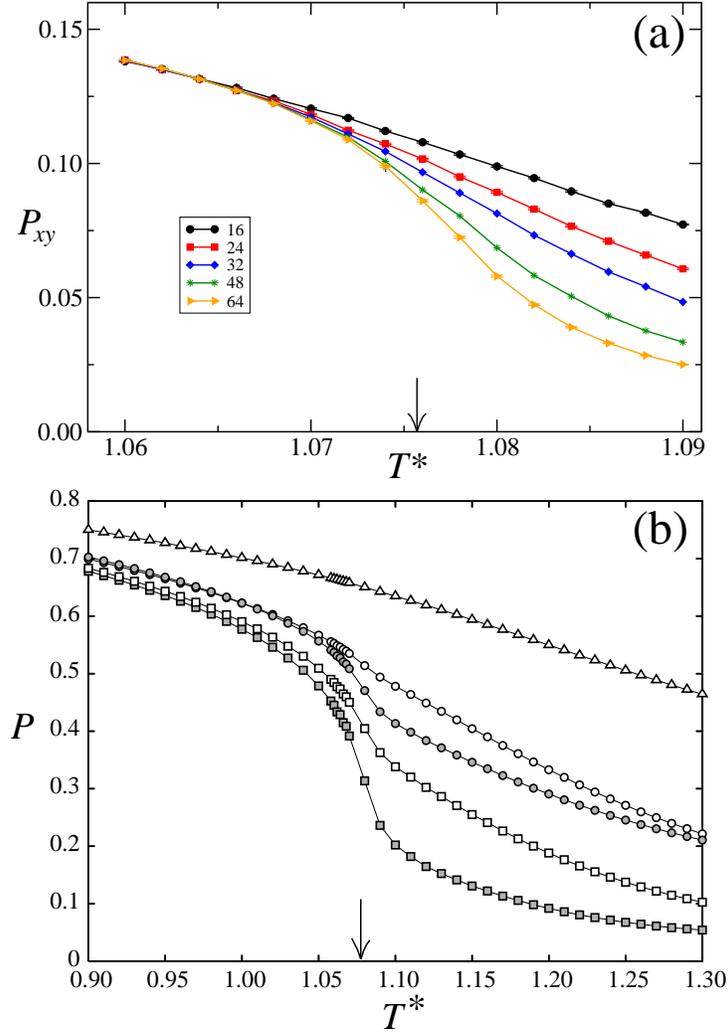}
\caption {\small{(Colour online). Order parameters (a) $P_{xy}$ and (b) $P$ 
as a function of temperature $T$, both for pore width
$h=8$. (a) Different curves give $P_{xy}$ for different values of lateral size $L$ (see key).
(b) Order parameter $P_i$ for $i=1$ (triangles), $i=2$ (open circles), $i=3$ (open squares)
and $i=4$ (filled squares) for lateral size $L=32$. 
The global order parameter $P$ is represented by filled circles.
In both (a) and (b) the vertical arrow indicates the location of the phase transition
as estimated from the heat capacity.}}
\label{fig.pxy}
\end{figure}

\begin{figure}
\centering
\includegraphics[height=8.0cm,angle=0 ]{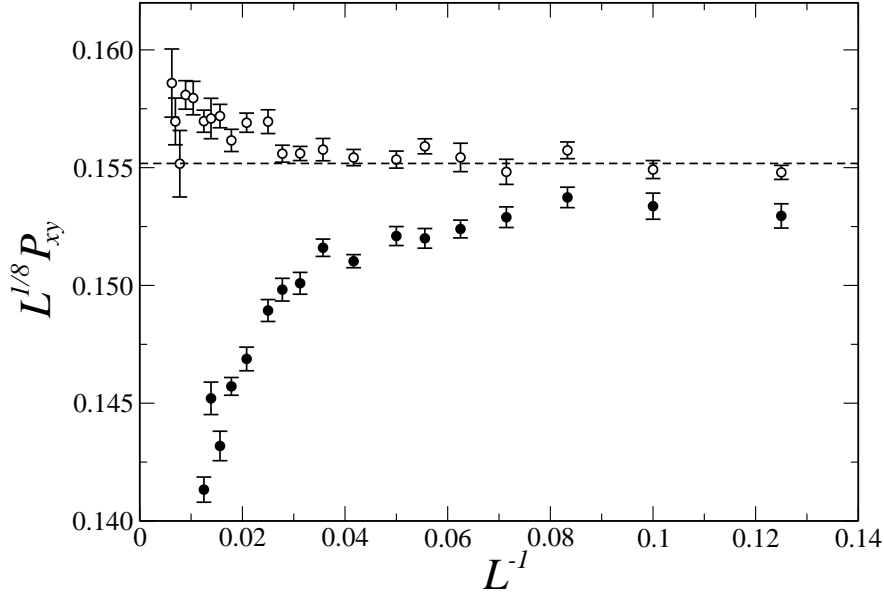}
\caption {\small{Dependence of the order parameter $P_{xy}$ on the lateral size $L$
of the system for two different values of scaled temperature which are close to the
true critical temperature, for the case $h=8$. Filled circles: 
$T^*=1.076$. Open circles: $T^*=1.075$. Error bars are included in each case.
The horizontal line indicates an approximate value of $L^{1/8}P_{xy}$ for the latter
temperature in the thermodynamic limit $L\to\infty$.}}
\label{pxys}
\end{figure}

%{\color{red} In Fig. \ref{fig.pxy} we present results for the quantities 
{In Fig. \ref{fig.pxy} we present results for the quantities 
$P_{xy}$, $P$, and $P_i$ as a function of temperature for the fixed pore width $h=8$.
In part (a) the dependence of $P_{xy}$ on lateral size $L$ is shown. 
We see that lateral size hardly affects the value of $P_{xy}$ in the L phase
(low temperatures), while the value in the S phase (high temperatures) 
decreases with lateral size (the location of the transition
is indicated by an arrow). There is no clear signature of the transition at the 
level of $P_{xy}$. To check whether $P_{xy}\to 0$ in the S phase in the
thermodynamic limit, we have performed extensive simulations for systems with 
rather large lateral size. The results are plotted in Fig. \ref{pxys}, which
represents $L^{1/8}P_{xy}$ as a function of $L^{-1}$ (the exponent $1/8$
corresponds to a two-dimensional Ising-like critical transition). From these
results one can conclude that the transition has a two-dimensional character,
at least for the pore size $h=8$ and smaller (the nature of the transition should
change to first order for sufficiently wide pores, see discussion in Sec. \ref{wide}).

The uniaxial order parameter, $P_i$,  is plotted in Fig. \ref{fig.pxy}(b) for a fixed
lateral size of $L=32$ and for the different planes $i=1,2,3$ and $4$ (planes
with $i=8,7,6$ and $5$ are symmetric). The global order parameter $P$ is also
plotted. At the transition (indicated by an arrow) the order parameter
shows a larger variation, but again no anomaly can be seen. Note that the variation
with temperature is more abrupt for the planes closer to the middle of the pore,
which is the region where the director is having more dramatic rearrangements.
As opposed to $P_{xy}$, the global uniaxial order parameter $P$ should be finite in the 
thermodynamic limit at the transition, and in this limit a kink should exist; again the 
finite lateral size prevents this anomaly to show up.

The picture that emerges from these results is that, starting from the low-temperature
region, where the L phase is stable, and on approaching the transition by increasing
the temperature, the director tilt angle starts to bend from the linear-like
configuration and ultimately develops an abrupt variation that becomes a step
at the transition (so that $(P_{xy})_i=0$ at each plane, implying
$\sin{2\phi_i}=0$). This conclusion is subtle, as it implies that the tilt-angle
profiles shown in Fig. \ref{fig.p}(d) for the critical configuration actually tend 
to a step-function in the thermodynamic limit $L\to\infty$.}

The physical nature of the phase transition is easily explained as a 
competition between the anchoring effect of the walls, which the film tries
to satisfy simultaneously but creates conflicting director orientations at the
two walls, and the elastic energy incurred when the director rotates between
one orientation and the other. At low temperatures or large film thickness,
the system can accommodate a linearly rotating director in the film.
When the temperature is high or the film thin, the system prefers to eliminate
the (large) elastic contribution at the cost of creating a step configuration, 
which can be regarded as a planar defect.

The L phase is degenerate in the following sense. As one goes 
from $z=1$ to $z=h$ through a line of sites with equal values 
of $x$ and $y$, the orientation of the spins rotates from 
$\hat{\bm x}$ to $\hat{\bm y}$. This rotation can be clockwise ($+$)
or anticlockwise ($-$). The NN interactions between sites impose correlations
between pairs of NN site lines, which make favorable that two NN lines have
the same rotation sign. Below $T_c$ the system 
chooses (with equal probability) either $+$ or $-$ as the preferred orientation
sign. % {\color{red}PERO ESTO SE APLICA IGUALMENTE A LA FASE S...!}

%Both the L and S phases are degenerate in the following sense. As one goes 
%from $z=1$ to $z=h$ through a line of sites with equal values of $x$ and $y$, the orientation
%of the spins rotates from $\hat{\bm x}$ to $\hat{\bf y}$. This rotation can be clockwise ($+$)
%or anticlockwise ($-$). The NN interactions between sites impose correlations
%between pairs of NN site lines, which make favorable that two NN lines have
%the same rotation sign. The system always chooses (with equal probability) one of the
%two signs. 

\subsection{h=1}

\begin{figure}
\centering
\includegraphics[height=7cm,angle=0 ]{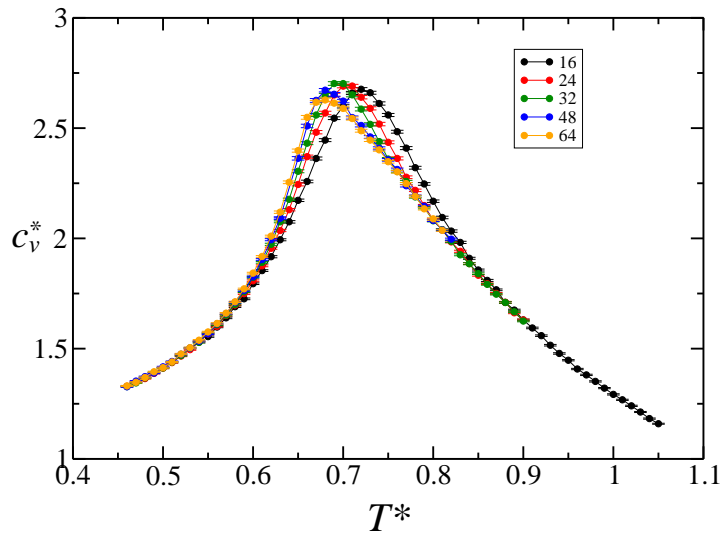}
\caption {\small{(Colour online). Variation of the excess heat capacity per particle $c_v^*$ with
reduced temperature $T^*$ for $h=1$ and different values of $L$ (indicated in the inbox).}}
\label{Cvh1}
\end{figure}

\begin{figure}
\centering
\includegraphics[height=7cm,angle=0 ]{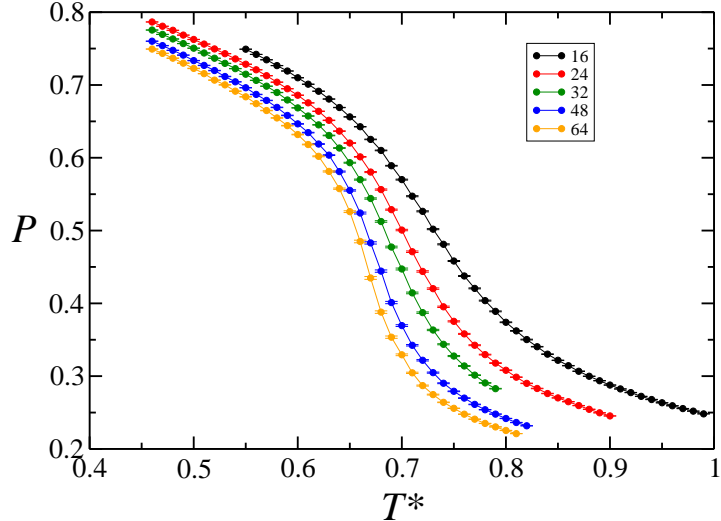}
\caption {\small{(Colour online). Behavior of the nematic order parameter $P$ as a function of reduced 
temperature $T^*$ for $h=1$ and various lateral sizes $L$ (indicated in the inbox).}}
\label{Ph1}
\end{figure}

The case $h=1$ (single layer) is special. Here the spins are subject to
an azimuthally-invariant potential that favours spin configurations
parallel to the plates. Therefore the transition belongs to the XY 
universality class. In fact, our results for
the heat capacity (see Fig. \ref{Cvh1}, and Table \ref{table.h1}) and the behavior of the nematic
order parameter (see Fig. \ref{Ph1}) suggest that the transition is of the
Berezinskii-Kosterlitz-Thouless (BKT) \cite{BKT1,BKT2}
type. The heat capacity shows a maximum, but $c_v^{\rm max}(L)$ hardly depends
on system size and presents a shift towards slightly lower
temperatures as $L$ increases. For a given system size $L$ we consider the temperature at which $|dP/dT|$ is maximum
as the corresponding pseudo-critical temperature $T_c(L)$.  With the values for
different system sizes a rough estimation of the transition temperature in
the thermodynamic limit, $T_{\hbox{\tiny BKT}}= \lim_{L\rightarrow \infty} T_c(L)$,
can be obtained by fitting the results to the equation \cite{Tomita}:
\begin{equation}
T_c(L) \simeq T_{\hbox{\tiny BKT}} + a \left( \log L \right)^{-2}.
\end{equation}
With this scheme we obtain $T_{\hbox{\tiny BKT}}^* = T_c^*(h=1)  \simeq 0.63 \pm 0.01$.

\begin{table}
\caption{\small{Maximum excess heat capacity per particle, and the corresponding temperatures $T_{\rm max}$,
and pseudo-critical temperatures, $T_c(L)$, defined as indicated in the text, for
the hybrid nematic film for pore width $h=1$.}}
\begin{center}
\begin{tabular}{@{}lllllllll}\hline\hline
\multicolumn{1}{c} {} &
\multicolumn{1}{c} {$L=16$} &
\multicolumn{1}{c} {$L=24$} &
\multicolumn{1}{c} {$L=32$} &
\multicolumn{1}{c} {$L=48$} &
\multicolumn{1}{c} {$L=64$} \\
\hline

$c_{v}^{\rm max}(L)/k $  & $2.661$(3) &  $2.668$(3) & $2.696$(4) & $2.628$(3)& $2.614$(3) \\
$T^*_{\rm max}(L) $  & $0.7210$(7) &  $0.7040$(5) & $0.6940$(7) & $0.6872$(4)& $0.6858$(4) \\
$T^*_c(L)$ & 0.723(1) & 0.703(1) & 0.691(1) & 0.678(1) & 0.669(1) \\
\hline\hline
\end{tabular}
\label{table.h1}
\end{center}
\end{table}

\subsection{Wide pores and global phase diagram}
\label{wide}

Using the techniques explained in the previous sections, we have extended
the calculation of the LS transition to other values of pore width $h$.
The values of $T_c(L,h)$ obtained from the heat capacity are used to extrapolate to the 
thermodynamic limit, using Eqn. (\ref{eq.tclh}). The results of this fitting for
the different values of $h$ explored are gathered in Table \ref{table.tch}.
As can be seen, the critical temperature $T_c(h)$ increases monotonically with $h$ and
approaches the value of the bulk isotropic-nematic transition temperature,
$T_{\rm IN}^*=1.1225(1)$ \cite{Priezjev,AML}. One important point is that only a single peak is 
observed in the specific heat in all cases as $T$ is varied, indicating the presence of a single
transition in this system. Therefore, our data do not corroborate the findings of
Chiccoli et al. \cite{Zann2}, who claim the existence of two distinct peaks in
the heat capacity.

\begin{table}
\begin{center}
\caption{Estimates of the transition temperatures for the hybrid
nematic films.
Error bars are given between parentheses, in units of the last figure
quoted, and correspond to 95\% confidence level. 
}
\begin{tabular}{cccccc}
\hline\hline
$h$  & 2&4&8&16 & 32 \\\hline
$T_c^*(h)$ & 0.817(1) & 0.994(2) & 1.076(1) & 1.108(1) & 1.119(1) \\
\hline\hline
\end{tabular}
\label{table.tch}
\end{center}
\end{table}

\begin{figure}
\includegraphics[width=5.2in]{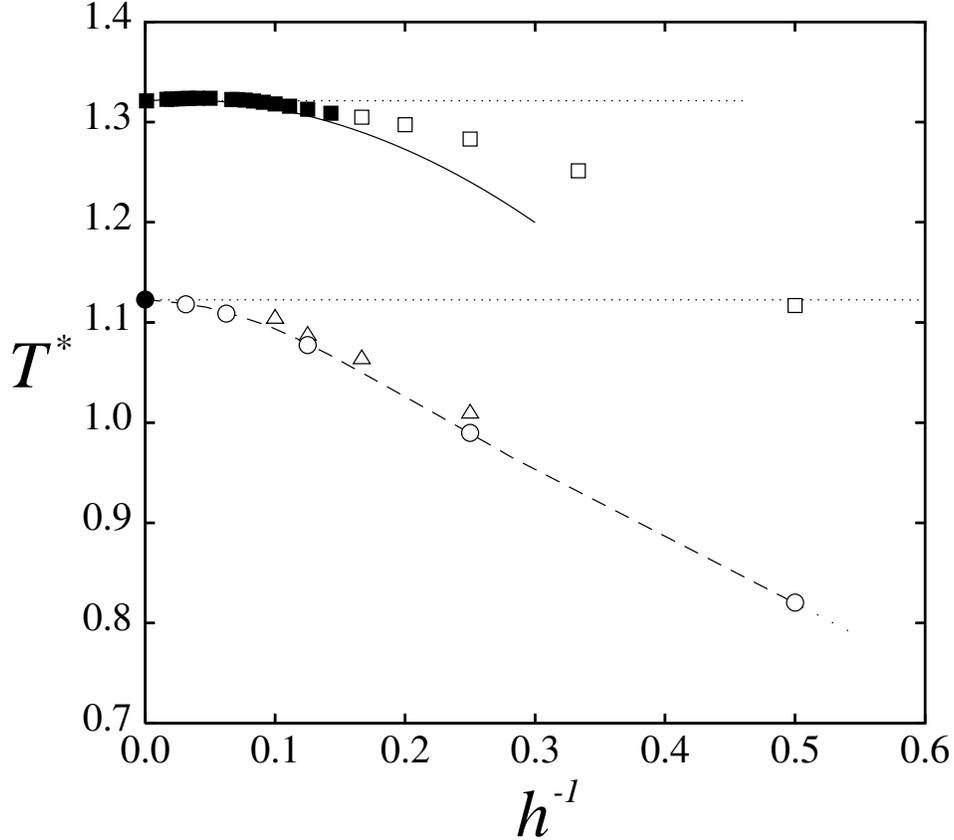}
\caption{Phase diagram for the hybrid cell in the $T^*$-$h^{-1}$
plane for the case {$\epsilon_s^{(1)}=\epsilon_s^{(2)}=\epsilon$}, 
showing temperatures at which the LS transition
occurs for each value of plate separation. 
L (S) phase is stable below (above) the
corresponding symbol. Circles: present MC simulation
results. Triangle: MC simulation results by Chiccoli et al. \cite{Zann1}.
Squares: present MF results. Filled symbols represent first-order
phase transitions, while open ones refer to continuous phase transitions.
Horizontal dotted lines: bulk temperatures as obtained from the MF
and MC calculations (upper and lower lines, respectively). 
Continuous line: modified Kelvin equation. Dashed line is a guide to the eye.}
\label{PDasym}\end{figure}

The resulting phase diagram in the plane $T$-$h^{-1}$ is presented in Fig.
\ref{PDasym}. The interval $0\le h^{-1}\le 1$ was covered in the MC simulations
(the bulk, $h^{-1}=0$, value was obtained from independent simulations in Ref. 
\cite{Priezjev,AML}). The maximum plate separation considered for the confined fluid
was $h=32$, which increases the maximum value used in \cite{Zann1}
and \cite{Zann2}. The LS transition line
spans the whole interval $0\le h^{-1}\le 1$. For the plate separations
explored, $1\le h\le 32$, the transition is continuous. As mentioned before, since the bulk
transition is of (weakly) first order, there must be a change from
first-order to continuous behaviour at some (probably large) value of $h$.

\begin{figure}
\includegraphics[width=17cm]{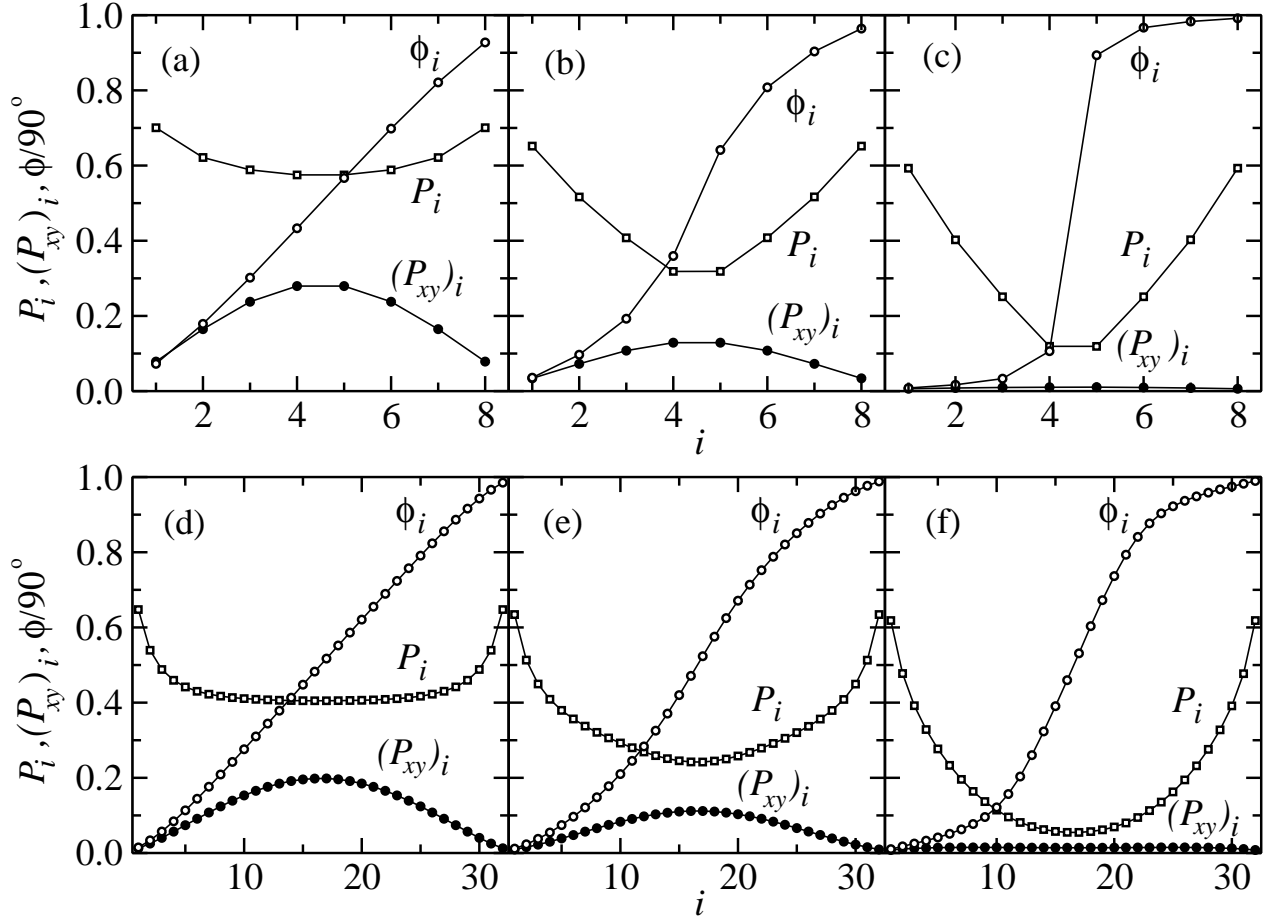}
\caption{Local order parameters $P_i$, $(P_{xy})_i$ and director tilt
angle $\phi_i$ obtained from the MC simulations for two different pore widths,
$h=8$ ($L=64$) and $h=32$ ($L=48$), at various temperatures in the neighbourhood of the
corresponding critical temperature $T_c(h)$. (a) $h=8$ and $T^*=1$;
(b) $h=8$ and $T^*=1.076$; (c) $h=8$ and $T^*=1.15$; (d) $h=32$ and $T^*=1.101$;
(e) $h=32$ and $T^*=1.118$; (f) $h=32$ and $T^*=1.135$.
Notice that for $T>T_c$ the jump in $\phi$ is system-size ($L$) dependent,
and becomes steeper as $L$ approaches the thermodynamic limit.}
\label{MCprof}\end{figure}

As the transition line is crossed at fixed $h$, the spin structure in
the slab changes suddenly but continuously, as it corresponds to the continuous
phase transition discussed in Sec. \ref{h8}. Here we show profiles for the cases 
$h=8$ and $h=32$, in order to illustrate the differences between narrow and wide pores.
Fig. \ref{MCprof} shows the change
in structure for the cases $h=8$ and $32$ as the temperature is increased, reflected
by the values of the order parameters $P_i$ and $(P_{xy})_i$, and by the director
tilt angle $\phi_i$. At high temperature [Figs. \ref{MCprof} (c) and (f)] the
structure is of the S type, 
with an abrupt change in the tilt angle as the middle plane of the slab is
crossed, and with a low value of the order parameter $P$ in the central region.
As $T$ is lowered, we pass from the S to the L structure, with the tilt angle
slowly rotating from one plate to the other. Note that the value of the
order parameter $P_i$ in the slab increases substantially at the transition [which
occurs in the situations represented by panels (b) and (e)]. In the S phase the difference between 
the two cases shown in the figure, which may be representative of a thin ($h=8$)
and a thick ($h=32$) slab, is that, in the thick-slab case, the nematic
films next to the plates are more separated, leaving a wider orientationally disordered
region in the central part of the slab. The reason why the central region of the pore
is not completely disordered, panel (f), may be a finite-size effect. Indeed, as discussed
in Sec. \ref{h8} (see Fig. \ref{pxys}), we expect a step-function behaviour for $\phi_i$ at the transition
[panel (e)] in the thermodynamic limit, 
while $P_i$ should go to zero right at the middle of the pore, and $(P_{xy})_i$ should
be zero everywhere. Therefore, in the situation described in panel (f), the tilt-angle
profile should be a step function, while the $P_i$ profile would be expected to exhibit
a wide gap with $P_i\simeq 0$, i.e. a thick isotropic central slab.  
%related to the presence of thick plate-nematic and
%isotropic-nematic interfaces, associated with the presumably long correlation length 
%in the model, as discussed later). 
As $h$ is increased, this central region will become thicker, implying that the S phase is
the confined phase connected with the {\it bulk} isotropic phase. On the low-$T$
side, the linear-like phase is the confined nematic phase and it
evolves to the {\it bulk} nematic phase (with a director that rotates 
more and more slowly across the slab). The LS transition is the
isotropic-to-nematic (IN) transition in a hybrid cell, and no additional 
capillary or structural transitions should be expected to occur in this system. 

{As a final comment, we note that the MC data for the transition points
obtained by Chiccoli et al. \cite{Zann1} are slightly shifted with respect
to our own data. These differences may result from the more efficient sampling
of the present study, which considers cluster algorithms in the MC moves. Also,
our simulations are longer, maximum lateral sizes are larger, and a proper 
finite-size calculation of the transition temperature is performed.} 

\section{Mean-field model}
\label{mean}

The MF theory for the Lebwohl-Lasher model has been used before to
study symmetric nematic slabs \cite{Todos1,Todos,Todos2}. A rich phase diagram
with respect to the parameters $T$, $h$ and surface couplings results.  
Here we use the model to rationalise the MC findings shown in the previous section,
focusing on the hybrid cell.
First we briefly comment on the implementation of the theory and then present the
results and their connection with the macroscopic behaviour.

\subsection{{Theory and method of solution}}

The orientational distribution of a spin in the $i$th plane is given by the
function $f_i(\hat{\bm s})$.
The complete MF free-energy functional for the Lebwohl-Lasher model is
\begin{eqnarray}
\hspace{-0.5cm}\frac{F[\{f_i\}]}{L^2}&=&
kT\sum_{i=1}^h\int d\hat{\bm s}f_{i}(\hat{\bm s})
\log{\left[4\pi f_{i}(\hat{\bm s})\right]}-
2\epsilon\sum_{i=1}^{h}
\int d\hat{\bm s}\int d\hat{\bm s}^{\prime}f_{i}(\hat{\bm s})
f_{i}(\hat{\bm s}^{\prime})P_2(\hat{\bm s}\cdot\hat{\bm s}^{\prime})
\nonumber\\\nonumber\\\hspace{-0.5cm}&-&
\epsilon\sum_{i=1}^{h-1}\int d\hat{\bm s}\int d\hat{\bm s}^{\prime}
f_{i}(\hat{\bm s})f_{i+1}(\hat{\bm s}^{\prime})
P_2(\hat{\bm s}\cdot\hat{\bm s}^{\prime})\nonumber\\\nonumber\\\hspace{-0.5cm}&-&
\epsilon_s^{(1)}\int d\hat{\bm s}f_{1}(\hat{\bm s})
P_2(\hat{\bm s}\cdot\hat{\bm m}_1)-
\epsilon_s^{(2)}\int d\hat{\bm s}f_{h}(\hat{\bm s})
P_2(\hat{\bm s}\cdot\hat{\bm m}_2)-\sum_{i=1}^h\lambda_i\int
d\hat{\bm s}f_i(\hat{\bm s}).
\end{eqnarray}
%where we defined {\color{red}$\epsilon^*=\epsilon/kT$ and 
%$\epsilon_s^{(i)*}=\epsilon_s^{(i)}/kT$}.
The $\lambda_i$'s are Lagrange multipliers ensuring the normalisation
$\int d\hat{\bm s}f_i(\hat{\bm s})=1$.
The interaction part contains contributions from spins on the same layer and
from spins on two neighbouring layers, and also from
the external potentials. Here we find it more convenient to use $\hat{\bm m}_1=\hat{\bm x}$ 
and $\hat{\bm m}_2=\hat{\bm z}$ as easy axes. 

Functional minimisation of $F$ provides the corresponding coupled, self-consistent Euler-Lagrange 
equations for each plane, which are projected onto a spherical-harmonics basis using
\begin{eqnarray}
f_i(\hat{\bm s})=\sum_{l=0}^{\infty}\sum_{m=-l}^{l}f_{lm}^{(i)}
Y_{lm}(\hat{\bm s}).
\label{expansion}
\end{eqnarray}
As usual in MF theory, the corresponding equations can be
interpreted as if each spin felt an effective field created by their neighbours. 
The effective field is given by the functions $\varPhi^{(\alpha)}(\hat{\bm s})$, with
\begin{eqnarray}
\varPhi^{(0)}(\hat{\bm s})=P_2(\cos{\theta}),\hspace{0.2cm}
\varPhi^{(1)}(\hat{\bm s})=
\sin{2\theta}\cos{\varphi},\hspace{0.2cm}
\varPhi^{(2)}(\hat{\bm s})=\sin^2{\theta}\cos{2\varphi},
\end{eqnarray}
and $(\theta,\varphi)$ the spherical angles of the spin $\hat{\bm s}$.
Instead of using the whole distribution functions $f_i(\hat{\bm s})$, 
the order will be described by three lab-fixed order parameters,
$\eta^{(\alpha)}_i$ where $\alpha=0,1,2,$ and $i=1,...,h$ runs through 
the $h$ layers. The order parameters are related to the $l=2$-subspace coefficients
$f_{lm}^{(i)}$ by $f_{20}^{(i)}=\eta_i^{(0)}\sqrt{5/4\pi}$,
$f_{21}^{(i)}=-f_{2,-1}^{(i)}=-\eta_i^{(1)}\sqrt{5/6\pi}$ and
$f_{22}^{(i)}=f_{2,-2}^{(i)}=\eta_i^{(2)}\sqrt{5/6\pi}$.
In terms of $\eta^{(\alpha)}_i$, the Euler-Lagrange equations are
written
\begin{eqnarray}
\eta^{(\alpha)}_i=\left<\varPhi^{(\alpha)}(\hat{\bm s})\right>_i,
\hspace{0.3cm}i=1,2,...,h,
\label{eqns1}
\end{eqnarray}
where $\left<...\right>_i$ are averages over the orientational
distribution function $f_i(\hat{\bm s})$, with
\begin{eqnarray}
f_i(\hat{\bm s})\propto e^{\displaystyle
\beta\epsilon\sum_{\alpha=0}^2\left(4\eta_i^{(\alpha)}
+\eta_{i-1}^{(\alpha)}+\eta_{i+1}^{(\alpha)}\right)
\varPhi^{(\alpha)}(\hat{\bm s})+\varPhi^{(s)}_i(\hat{\bm s})},
\label{eqns2}
\end{eqnarray}
where $\beta=1/kT$.
In this expression $f_i(\hat{\bm s})$ has to be normalised to unity, and
we take $\eta_{0}^{(0)}=-1/2$, $\eta_{0}^{(1)}=0$, $\eta_{0}^{(2)}=1$, $\eta_{h+1}^{(0)}=1$,
$\eta_{h+1}^{(1)}=0$ and $\eta_{h+1}^{(2)}=0$.
$\varPhi^{(s)}_i(\hat{\bm s})$ are surface fields, with the properties
\begin{eqnarray}
\varPhi^{(s)}_i(\hat{\bm s})=\left\{\begin{array}{ll}
\displaystyle -\frac{\epsilon_s^{(1)}}{2}\left[\varPhi^{(0)}(\hat{\bm s})-
\frac{3}{2}\varPhi^{(2)}(\hat{\bm s})\right],&i=1,\\\\
0,&1<i<h,\\\\
\epsilon_s^{(2)}\varPhi^{(0)}(\hat{\bm s}),&i=h.\end{array}\right.
\end{eqnarray}
The order parameters $\eta_i^{(\alpha)}$ are related to the eigenvalues of
the order tensor, $P_i$ (uniaxial) and $B_i$ (biaxial) order 
parameters, and the director tilt angle $\phi_i$, through the relations
\begin{eqnarray*}
&&\eta_i^{(0)}=P_iP_2(\cos{\phi_i})+\frac{3}{4}B_i\sin^2{\phi_i},\\\\
&&\eta_i^{(1)}=\left(\eta_i^{(0)}-\frac{\eta_i^{(2)}}{2}\right)\tan{2\phi_i},\\\\
&&\eta_i^{(2)}=P_i\sin^2{\phi_i}+\frac{1}{2}B_i\left(1+\cos^2{\phi_i}\right).
\end{eqnarray*}
Here $\phi_i$, for the sake of convenience, is measured with respect
to the $z$ axis (we remind the reader that, due to the symmetry of the model, this
angle is the same as the one used in the MC simulations). From these equations, we 
can obtain $P_i$, $B_i$ and $\phi_i$ from $\eta_i^{(0)}$, $\eta_i^{(1)}$ and $\eta_i^{(2)}$.
For the bulk system the surface fields are eliminated and $\eta_i^{(0)}=P$,
$\eta_i^{(1)}=\eta_i^{(2)}=0$. The isotropic-nematic phase transition is of first order,
and occurs at $T^*=1.321$. The order parameter at the transition is $P=0.429$.

\begin{figure}
\includegraphics[width=5.2in]{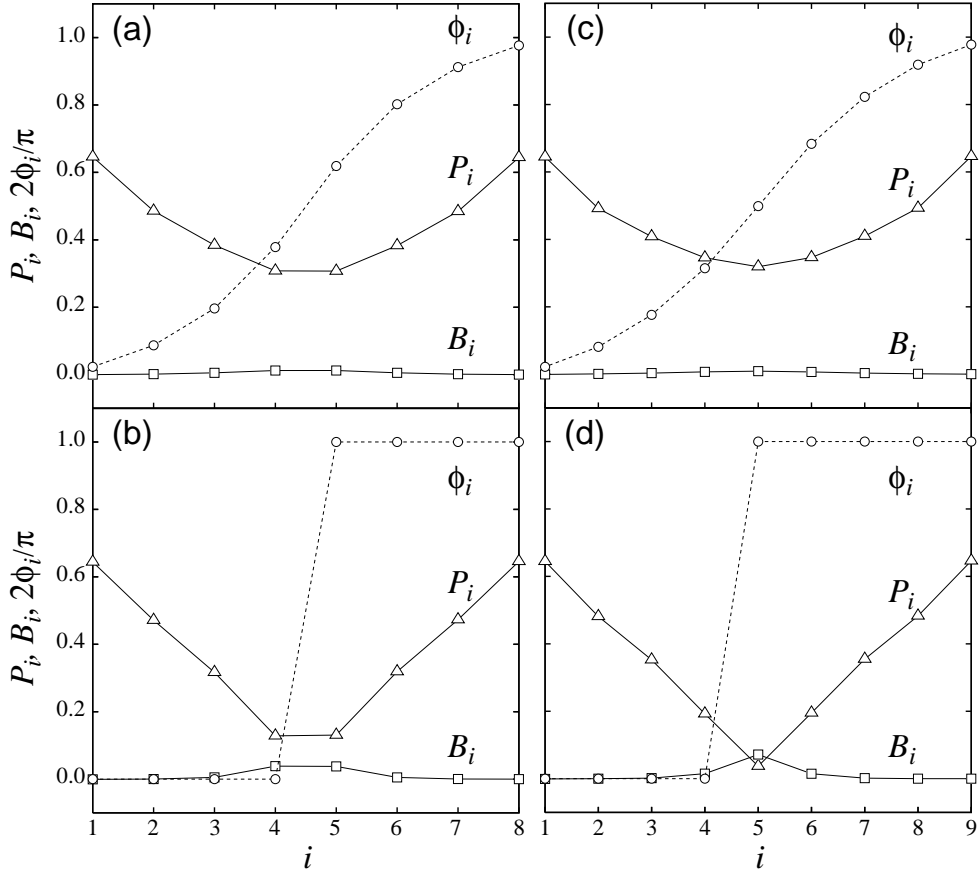}
\caption{Order parameters $P_i$ and $B_i$, and director tilt angle $\phi_i$ for the two 
phases coexisting at the LS phase transition, in the case
$\epsilon_s^{(1)}=\epsilon_s^{(2)}=\epsilon$ and as obtained from MF calculations. 
(a) L phase for $h=8$; (b) S phase
for $h=8$; (c) L phase for $h=9$; (d) S phase for $h=9$.}
\label{profMF}\end{figure}

\subsection{{Results: identical surface couplings}}

{In this section we consider the confined case and take 
$\epsilon_s^{(1)}=\epsilon_s^{(2)}=\epsilon$. These values ensure that, at bulk
conditions, both surfaces are wet by the nematic phase (see Appendix \ref{Wett}) so that,
close to the bulk transition temperature $T_{\hbox{\tiny IN}}$, thick nematic films are
expected at both surfaces.}

Order-parameter and tilt-angle profiles are shown in Fig. \ref{profMF} for the cases
$h=8$ and $9$ at the corresponding transition temperatures. The L and S structures
coexist at a first-order phase transition, in contrast to the MC results, which 
indicate a continuous transition.
As in the case of the MC results deep into the S phase, we note the clear 
discontinuity in the director tilt angle in the coexisting S phase. In the coexisting L phase
the director configuration adopts a linear-like configuration.
Also note that, at the transition, the nematic order parameter $P$ changes
quite substantially: in the S phase two nematic slabs meet at the central
region, such that the central spins are almost completely disordered, whereas the
L phase corresponds to a well-developed nematic slab. The differences between the cases 
where $h$ is an even or odd number are apparent by comparing the cases $h=8$ and $h=9$.
While the L phase hardly changes, the S phase of the even-$h$ case does not have a 
negligible value of the order parameter $P$ at the central region, in
contrast with the midpoint of the odd-$h$ slab. {The biaxial order parameter $B$ is 
non-negligible only in the neighbourhood of the step}. The uniaxial order parameter and director tilt
angle profiles obtained from the MF theory are quite similar to those from
MC simulation [cf. the two coexisting phases of Figs. \ref{profMF} (a) and (b) with
the structures shown in Figs. \ref{MCprof}(a) and (c)].

\begin{figure}[h]
\includegraphics[width=3.2in]{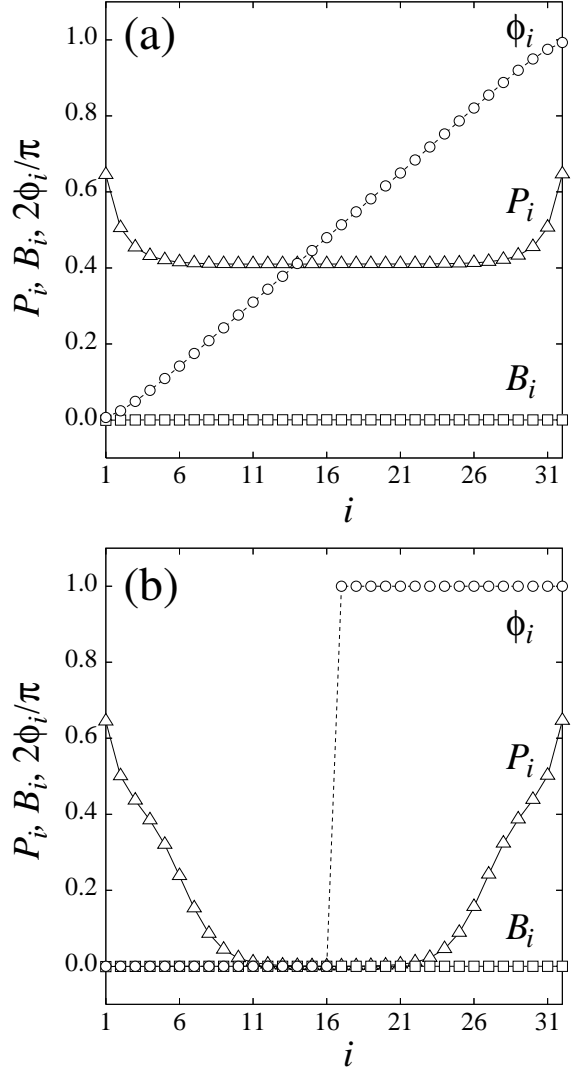}
\caption{Order parameters $P_i$ and $B_i$, and director tilt angle $\phi_i$ for the two 
phases coexisting for a pore width $h=32$, as obtained from the MF calculations,
in the case $\epsilon_s^{(1)}=\epsilon_s^{(2)}=\epsilon$.
(a) L phase. (b) S phase.}
\label{MF}\end{figure}

The nature of the LS transition 
becomes evident if we look at a wider pore. This is shown in Fig. \ref{MF}, which 
corresponds to $h=32$. In this case the high-temperature S phase has a large central region 
with a virtually zero value of $P$. As the transition is crossed from the region of high 
temperature, the value of $P$ in the central region increases to a nematic-like value, and
the total order parameter in the cell undergoes a discontinuous change. Therefore this transition, 
which corresponds to the capillary isotropic-nematic transition, is the same as the LS transition,
and it can be concluded that there is a single transition line in the phase diagram. 
Note that the phase corresponding to Fig. \ref{MF}(b) (confined isotropic phase, with
two differently-oriented nematic slabs adsorbed at each plate) for $h=32$
is smoothly connected to that of Figs. \ref{profMF}(b) for $h=8$ or 
(d) for $h=9$ (the step-like phase),
since they are actually the same phase but with an `isotropic' central slab of different width.

The MF phase diagram, in the plane $T$ vs.
$h^{-1}$, is presented in Fig. \ref{PDasym}. Transition temperatures for
the different values of $h$ explored are represented by squares.
Note that the character of the
transition changes from first order (for $h\ge 7$) to continuous (for $h\le 6$).
The main differences between the MF results and the MC simulations are: (i) the
transition is weakly of first order in MF for $h\ge 7$; in the simulations it is continuous for
the range of plate separations explored (as mentioned already,
the transition must change to first order at 
some, probably large, value of $h$, since it is of first order in bulk).
(ii) There is a shift in the transition to higher values 
of $T$ in MF, in correspondence with the shift in the bulk 
transition. (iii) In the simulation, the transition line seems to tend to the bulk value from below, 
with a very small slope at the origin $h^{-1}=0$ [see Fig. \ref{PDasym}];
in the MF theory, it changes slope and actually crosses the bulk temperature at $h\simeq 12$
(see Fig. \ref{PDasym1}, where an enlarged phase diagram is presented).

\begin{figure}[h]
\includegraphics[width=3.5in]{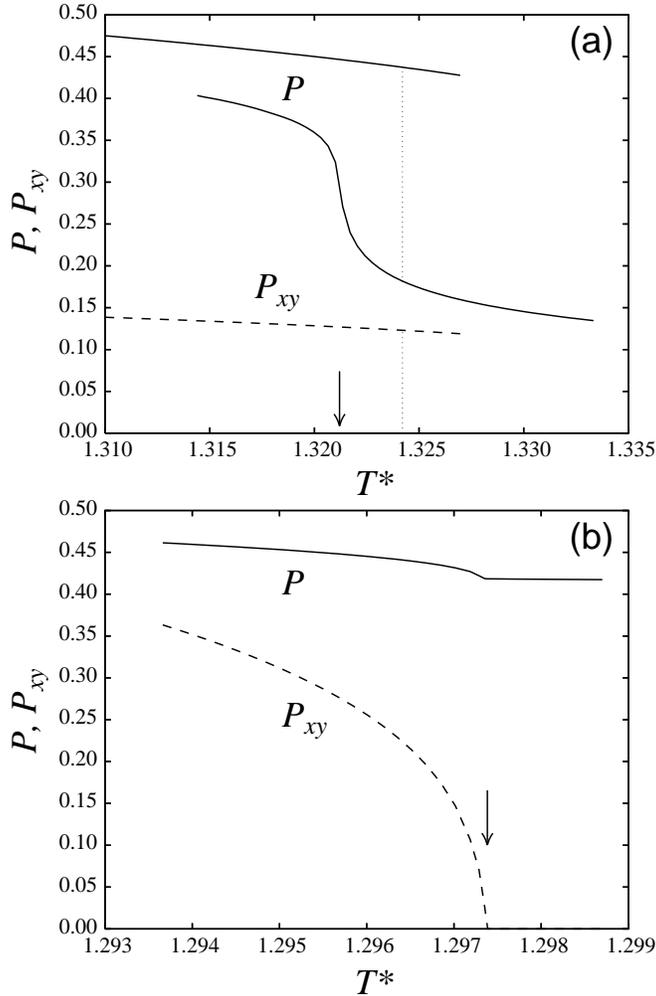}
\caption{Order parameters $P$ and $P_{xy}$ as a function of reduced temperature $T^*$ from mean-field 
theory. (a) $h=60$; (b) $h=5$. Continuous curves: $P$. Dashed curves: $P_{xy}$. In (a), vertical
dotted lines indicate location of first-order LS phase transition, while arrow points to bulk temperature.
In (b), arrow indicates temperature of continuous LS transition.}
\label{PPxy}\end{figure}

In Fig. \ref{PPxy} we plot the order parameters $P$ and $P_{xy}$ as a function of reduced
temperature $T^*$ for the case $h=60$ [panel (a)], where the LS transition is of first order, and
$h=5$ [panel (b)], where the transition is continuous. The behaviour of the order parameters
reflected in the figures may be qualitatively similar to the real situation. In (a), both order
parameters undergo discontinuous changes, indicated by the dotted vertical lines (the sharp
variation in $P$ in the metastable step-like branch below the transition corresponds to the
frustrated wetting transition at each plate {due to the confinement}). 
In panel (b) both order parameters are continuous
but exhibit a `kink' at the LS transition. Note that $P_{xy}$ is always zero in the step-like phase,
implying that the director tilt-angle is a perfect step function.

\begin{figure}
\includegraphics[width=5.2in]{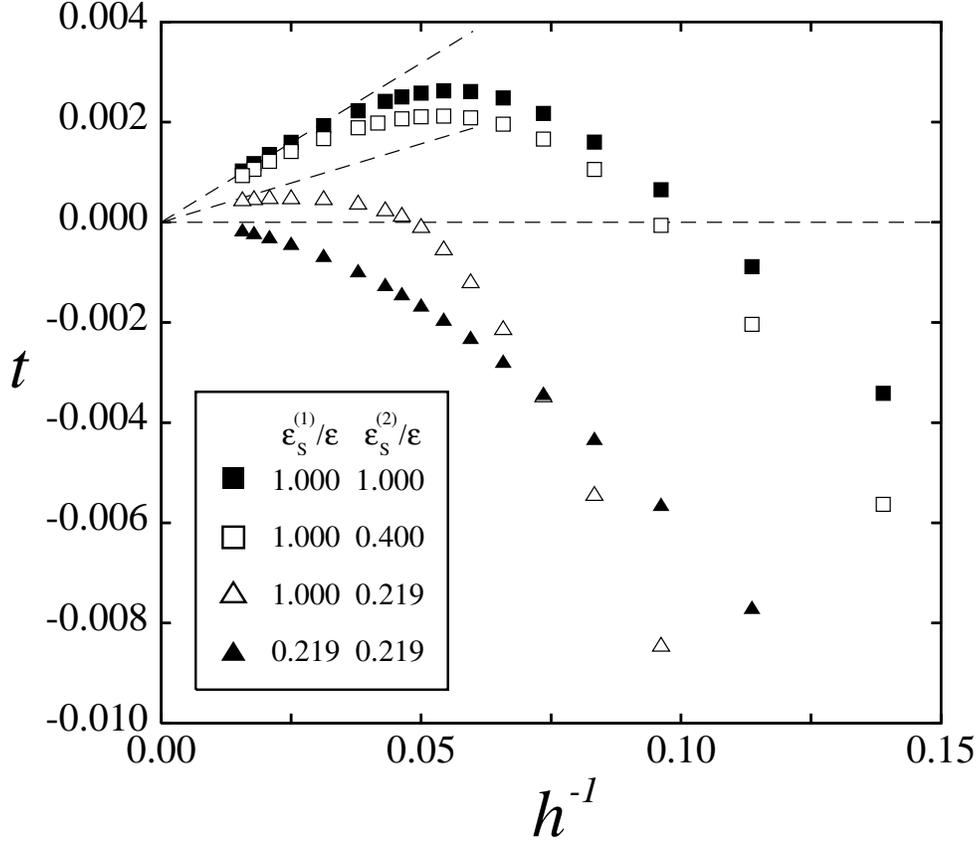}
\caption{Phase diagram for the hybrid cell in the $t$-$h^{-1}$ plane, with
$t=(T-T_{\hbox{\tiny IN}})/T_{\hbox{\tiny IN}}$, for different values of the surface
couplings $\epsilon_s^{(1)}$ and $\epsilon_s^{(2)}$ (indicated in the inbox), as
obtained from mean-field theory.
Dashed lines correspond to the lowest-order Kelvin equation in each case (see text).}
%dashed line corresponds to simple Kelvin equation without elastic contribution.
\label{PDasym1}\end{figure}

\subsection{{Results: other surface couplings}}

{We now discuss the case where the surface coupling constants are
different. This situation is closer to the experiments.
We analyse cases where the couplings of the two surfaces are different, and also
consider situations where conditions of complete wetting by nematic prevail, as well
as cases where one or the two surfaces are not wet by the nematic phase. 

\begin{figure}
\includegraphics[width=6.5in]{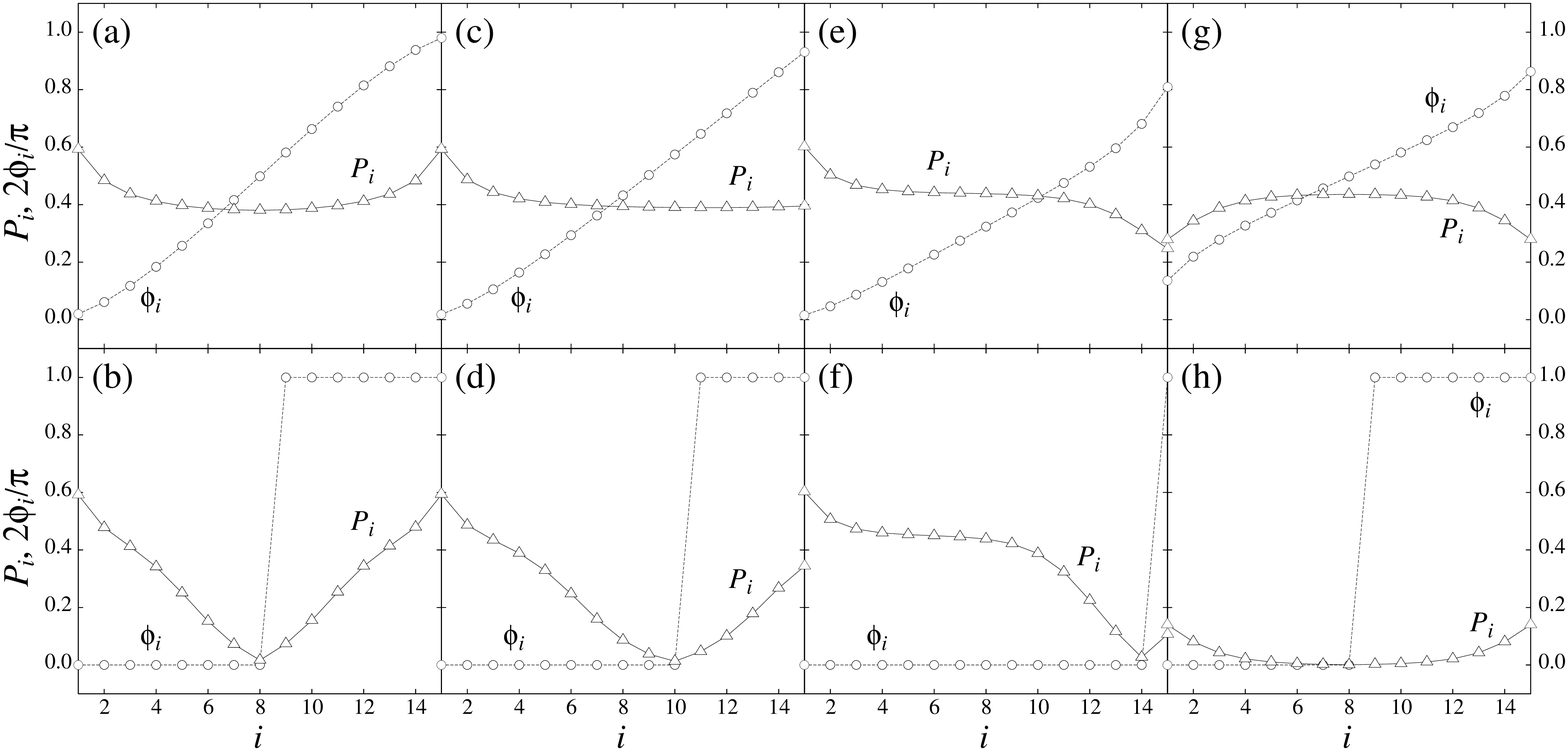}
\caption{Order parameter $P_i$ and director tilt angle $\phi_i$ for the two
phases coexisting at the LS transition for a pore of width $h=15$ and different
values of the surface coupling constants, as obtained from mean-field theory. Upper panels:
linear-like phase. Lower panels: step-like phase. The surface parameters 
$(\epsilon_s^{(1)},\epsilon_s^{(2)})$ are as follows: (a) and (b), 
$(\epsilon,\epsilon)$; (c) and (d), $(\epsilon,0.4\epsilon)$; (e) and (f), 
$(\epsilon,0.219\epsilon)$; and (g) and (h), $(0.219\epsilon,0.219\epsilon)$.}
\label{profs}
\end{figure}

\begin{itemize}
\item In the first case, the surface couplings are chosen as
$(\epsilon_s^{(1)},\epsilon_s^{(2)})=(\epsilon,0.4\epsilon)$, which again ensures a regime 
of complete wetting by the nematic phase at the two surfaces (see Appendix \ref{Wett}). 
Therefore thick nematic films are expected at both surfaces for temperatures close to the bulk 
transition temperature. In this case the LS phase-transition curve shifts to
lower temperatures with respect to the previous case, but by a small amount, as evident from 
Fig. \ref{PDasym1}. From a structural point of view, the change involves a shift in the location 
of the step: now it is not symmetrically located with respect to the two surfaces, but closer to the 
surface with the weakest coupling constant (i.e. the right surface). 
This feature can be seen in Figs. \ref{profs}(c) and (d), where
the uniaxial order-parameter profile $P_i$ and director tilt-angle $\phi_i$
are plotted for the case $h=15$ and for the two phases coexisting at the LS
transition. For comparison, the corresponding symmetric profiles for
the case $\epsilon_s^{(1)}=\epsilon_s^{(2)}=\epsilon$ are also plotted in panels
(a) and (b).

\item In the second case, the surface couplings 
are $(\epsilon_s^{(1)},\epsilon_s^{(2)})=(\epsilon,0.219\epsilon)$. Here conditions of
nematic wetting only prevail at one surface (Appendix \ref{Wett}). The shift in the
LS transition curve is much more drastic: the maximum is lower, and 
the curve crosses the bulk transition temperature at a higher value of $h$
(see Fig. \ref{PDasym1}).
For narrow pores, the profiles now reveal that the step is located next to the
weaker surface, as expected [see Figs. \ref{profs}(e) and (f)]. 
The pore width at which the transition 
changes from first to second order also moves to higher values (not shown in
Fig. \ref{PDasym1}).

\item Finally, we have examined the case 
$(\epsilon_s^{(1)},\epsilon_s^{(2)})=(0.219\epsilon,0.219\epsilon)$. Now
partial wetting applies in both surfaces and, as seen in Fig. \ref{PDasym1},
the LS transition curve exhibits no maximum.
The isotropic film in the slab centre is very wide and the two nematic films are not
in contact except for very narrow pores. The coexistence profiles for the linear- and
step-like phases for a pore of width $h=15$ are plotted in Fig. \ref{profs}(g) and (h).
\end{itemize}

In summary, as the surface coupling of one of the surfaces is made weaker, the step
moves towards that surface, and the LS transition temperature shifts to lower values. 
The cell width where the transition changes from first-order to continuous decreases.
The maximum in the curve also moves to higher values of pore width, and eventually
disappears. This feature is related to the wetting properties of the cell, as shown
in the macroscopic analysis of the following section.
}

\subsection{Connection with macroscopic behaviour}
\label{macro}

{Much of the behaviour shown in the previous sections can be explained 
using a simple macroscopic approach.} 
{For a fluid confined into a pore of width $h$, the macroscopic Kelvin equation 
gives the undercooling (or overheating) of the transition, with respect to the bulk transition, 
as $\Delta T(h)=T_c(h)-T_{\rm IN}=a_1h^{-1}$, with $h\to\infty$.
{As discussed in Appendix \ref{apC}, the coefficient $a_1$ can be related to 
the coexistence parameters $\Delta\gamma$ and $s_{\hbox{\tiny N}}$ as 
$a_1=\Delta\gamma/s_{\hbox{\tiny N}}$, where
$s_{\hbox{\tiny N}}$ is the nematic entropy density at the bulk IN transition, and 
$\Delta\gamma\equiv
\gamma_{\hbox{\tiny SI}}^{(1)}-\gamma_{\hbox{\tiny SN}}^{(1)}+
\gamma_{\hbox{\tiny SI}}^{(2)}-\gamma_{\hbox{\tiny SN}}^{(2)}$,
with the superscript denoting the type of
substrate, i.e. the left or right substrate (note that the value of the surface tensions
does not depend on the preferred surface orientation --as long as the director remains uniform--
but only on the value of the surface coupling $\epsilon_s^{(i)}$). 
For an isolated surface of type $i$ in contact with a bulk phase, the relation 
$-\gamma_{\hbox{\tiny IN}}\le\gamma_{\hbox{\tiny SI}}^{(i)}-\gamma_{\hbox{\tiny SN}}^{(i)}
\le\gamma_{\hbox{\tiny IN}}$ holds; the right equality corresponds to
wetting by nematic, while that in the left pertains to wetting by isotropic.
Therefore we may have $a_1<0$ or $a_1>0$,
and the transition curve $\Delta T(h)$ will monotonically decrease or increase with $h^{-1}$,
respectively, in the regime of large $h$. The sign of $\Delta\gamma$ depends on the
surface couplings: the difference  
$\gamma_{\hbox{\tiny SI}}^{(i)}-\gamma_{\hbox{\tiny SN}}^{(i)}$ vanishes for
$\epsilon_s^{(i)}=0.219\epsilon$, being positive (negative) for larger (lower) $\epsilon_s^{(i)}$.
For the different surface couplings analysed above, we have: 

\begin{itemize}
\item[(i)] $(\epsilon_s^{(1)},\epsilon_s^{(2)})=(\epsilon,\epsilon)$ and,
$(\epsilon,0.4\epsilon)$. Since nematic wetting occurs at both surfaces,
$\gamma_{\hbox{\tiny SI}}=\gamma_{\hbox{\tiny SN}}+
\gamma_{\hbox{\tiny IN}}$ --see Appendix \ref{Wett}, so that
$\Delta\gamma=2\gamma_{\hbox{\tiny IN}}=0.0351 kTa^{-2}$. The corresponding $a_1$ coefficient 
gives the dashed straight line plotted in Fig. \ref{PDasym1}; as can be seen, the data follow 
the behaviour predicted by the macroscopic analysis for large $h$.

\item[(ii)] $(\epsilon_s^{(1)},\epsilon_s^{(2)})=(\epsilon,0.219\epsilon)$.
Now wetting occurs only at one surface. Since 
$\gamma_{\hbox{\tiny SI}}^{(1)}=\gamma_{\hbox{\tiny SN}}^{(1)}$, we have
$\Delta\gamma=\gamma_{\hbox{\tiny IN}}=0.0176 kT$. Again the MF data follow
this behaviour in the regime of large $h$ (Fig. \ref{PDasym1}).

\item[(iii)] $(\epsilon_s^{(1)},\epsilon_s^{(2)})=(0.219\epsilon,0.219\epsilon)$.
Now partial wetting applies and $\Delta\gamma=0$. The LS transition curve 
departs horizontally from the $h$ axis and therefore exhibits no maximum, 
as indeed shown by the MF results in Fig. \ref{PDasym1}. 
\end{itemize}

As commented above, the MF results indicate that the two surfaces are wet by the nematic 
phase in the case $(\epsilon_s^{(1)},\epsilon_s^{(2)})=(\epsilon,\epsilon)$.
Whether this is also true in the MC simulations of Section \ref{wide} is not known, and a more 
detailed study of the wetting scenario would be necessary. Our present MC 
data seem to indicate $a_1<0$, which
would be incompatible with nematic wetting and even with preferential nematic adsorption, i.e.
$\gamma_{\hbox{\tiny SN}}<\gamma_{\hbox{\tiny SI}}$.
However, since the MC profiles indicate that the plates seem to adsorb preferentially the 
nematic phase, we could still have
$a_1>0$ in the real system, but with a change of regime at very large values of $h$.
Another factor to bear in mind is the presumably large correlation length of the model,
associated with the weakness of the bulk transition, which
would give rise to slowly decaying interfaces and to
the inapplicability of the Kelvin equation except for extremely wide pores. 
}

For smaller separations, the elastic effects in the linear-like phase must be very important, 
since the director rotates essentially between $0^{\circ}$ and $90^{\circ}$ in a very short
distance. These effects can be shown (Appendix \ref{apC}) to give rise to an additional
contribution to the Kelvin equation, namely \cite{Zann1}
$\Delta T(h)=a_1h^{-1}+a_2h^{-2}$, which is the so-called {\it modified Kelvin 
equation}. The first term comes from capillary forces, already discussed, whereas 
the second is due to elastic effects. The elastic contribution is always negative ($a_2<0$), 
promoting capillary isotropisation, and dominates the physics in the regime of narrow pores. 
It explains the decreasing behaviour of the LS transition curve for narrow pores.
The sign of the first term dominates for very wide pores, and if positive promotes
capillary nematisation, giving rise to a maximum in the LS transition curve when combined
with the second term.

In order to estimate the value of $a_2$, it is necessary to compute the elastic constants
of the model (Appendix \ref{elastic}). As shown in Appendix \ref{apC}, 
$a_2=K\pi^2/8s_{\hbox{\tiny N}}=-1.630a^2\epsilon k^{-1}$, with $K$ the model elastic constant. 
Fig. \ref{PDasym} compares the MF and macroscopic models (note that the comparison can only be
made in the regime where the transition is of first order). The overall agreement is 
not very good. For large $h$ the surface behaviour correctly predicts the capillary LS transition
(dashed lines in Fig. \ref{PDasym1}) but, {as soon as the pore becomes narrower, the elastic contribution 
comes in. However, because the transition is weakly first-order, the correlation
length, and therefore the interfacial thickness, is very large. Consequently,
in the S phase there are thick nematic layers at the walls, which violate the
assumptions of the model. In the other cases studied the agreement is also disappointing, in particular
in the partial-wetting cases.

\section{{Discussion}}
\label{discussion}

{The response of the system to confinement is intimately connected to
its wetting behaviour. This is especially important in connection with the observation of
the step-like structure. In a situation of complete wetting of the two surfaces
by the nematic phase, the thickness of the nematic films at temperatures close to
the clearing temperature $T_{\hbox{\tiny IN}}$ will be large, and the two films
with oposing directors will meet at the slab centre when $h$ is small, producing a
step-like phase which will turn into the linear-like phase as temperature is lowered.
As the pore gets wider, the step-like phase becomes the isotropic phase
with a nematic film adsorbed at each surface. In a partial-wetting situation, the nematic 
film thickness will be very small, the central isotropic region will be wide, and the two 
nematic films will never meet, except maybe for very narrow pores. 

In the light of our simulation results and the interpretation obtained from the MF theory,
it is interesting to discuss the quantity $h_{\rm max}$ introduced 
by Chiccoli et al., which these authors obtain from the intersection between their 
linearly-extrapolated data for the transition temperatures and the bulk temperature 
$T_{\hbox{\tiny IN}}$ (see Fig. 4 in \cite{Zann1}), i.e. the intersection between
a linear fit to the triangles in our Fig. \ref{PDasym} and the horizontal line 
$T^*=1.1225$. Our present MC results indicate that the LS transition curve is below
the bulk temperature, at least for the pore widths explored, and that the transition
continues as the confined IN transition up to $h=\infty$. Therefore,
the value $h_{\rm max}=16.6$ obtained by Chiccoli et al. from the extrapolated
data, and identified as the maximum slab thickness for which the structural phase 
transition can be found, is somewhat misleading,
as it seems to imply that this point terminates a phase transition curve; however, the
phase transition continues up to $h=\infty$, the step-like phase for narrow pores 
being smoothly connected (from a thermodynamic viewpoint)
with the confined isotropic phase for wider pores and eventually with the bulk isotropic 
phase for $h=\infty$. Our results imply that $h_{\rm max}$ obtained by Chiccoli et al.
does not seem to have any special meaning \cite{us}.
}

\section{Conclusions}
\label{conclusions}
In this paper we have studied the Lebwohl-Lasher model 
in a confined slit pore, using MC computer simulation and
MF theory. Two types of surface conditions have been imposed, namely
symmetric and asymmetric walls, with special emphasis on
the latter. The simulations and the data analysis have been carefully performed, with
a view to locating accurately the phase transition. For the symmetric walls, we have set 
a lower limit for the pore width at which the capillary isotropic-nematic transition takes 
place: the transition is still absent for $h=24$, but the behaviour of the heat 
capacity indicates that it might occur for slightly larger pore widths. 

The asymmetric slab was the central target of our investigations, and
consequently was studied in more detail. A phase transition, spanning the whole range in
pore widths and associated with a change from the linear-like to the step-like director
configurations (LS transition), was measured. 
In all the cases examined, $1\le h\le 32$, the transition was found to be continuous. 
The LS transition involves a structural change of the director configuration
but, from examination of the order-parameter profiles, it is evident that the transition
corresponds to the IN transition in a confined geometry in a situation where the slab
is subject to two conflicting favoured directions at the two surfaces. Therefore,
there is a single phase transition in the confined slab. These results are supported by
a MF theory, which gives qualitatively similar results. 
Even though the LS or IN transition is continuous in the range $1\le h\le 32$ according to the
simulations, the bulk case, $h=\infty$, presents a weakly first-order transition
(as obtained from independent simulations). This implies that there must be a change in order at some, probably large,
value of $h$; the MF model predicts $h=6$ but this value is clearly too small.
The case $h=1$ has also
been examined by MC simulation.  Contrary to the transitions in the case $1<h \le 32 $, which 
belong to the 2D Ising universality class, when $h=1$ the transition is essentially
different and pertains to the XY-model class.

{The case of different surface coupling constants was also analysed, using only MF theory.
The results are qualitatively similar, as long as the nematic phase wets both surfaces.
In this case the IN transition can be more clearly identified with the structural
transition studied in the literature. When partial wetting applies to one of the surfaces
the IN transition occurs between two phases, one of which is the linear-like phase;
the other, step-like phase, changes in this case to a phase with a director which is uniform in most of 
the slab volume. When neither surface is wet, the latter phase consists of a thick
central isotropic slab with thin nematic films on the two surfaces.
}

\acknowledgements
The authors gratefully acknowledge the support from the Direcci\'on General de Investigaci\'on
Cient\'{\i}fica y T\'ecnica under Grants Nos. MAT2007-65711-C04-04, MOSAICO, FIS2007-65869-C03-01, 
FIS2008-05865-C02-02, FIS2010-22047-C05-01 and FIS2010-22047-C05-04, and from the
Direcci\'on General de Universidades e Investigaci\'on de la Comunidad
de Madrid under Grant No. S2009/ESP-1691 and Program MODELICO-CM.
RGM would like to thank grant MOSSNOHO for a research contract.

\appendix 

\section{SURFACE TENSIONS AND WETTING PROPERTIES}
\label{Wett}

The surface tensions of the three interfaces involved are necessary to discuss
the wetting properties of the model and to investigate how the macroscopic behaviour is
obtained in the confined slab as $h\to\infty$. The isotropic-nematic interface $\gamma_{\hbox{\tiny IN}}$
was computed in slab geometry, by considering a slab of nematic material
sandwiched between two isotropic regions at the coexistence temperature $T^*=1.3212$. 
The uniaxial nematic order-parameter
profile is depicted in Fig. \ref{OPprofIN}. The surface tension obtained is
$\gamma_{\hbox{\tiny IN}}=0.0176 kT a^{-2}$. From a fit of the profile to 
a hyperbolic-tangent function, we get an interfacial
width (correlation length) of $\xi=2.63a\sim 3a$. We note that 
$\gamma_{\hbox{\tiny IN}}$ is relatively small and $\xi$ relatively large,
confirming the weak character of the bulk isotropic-nematic transition.

\begin{figure}[h]
\includegraphics[width=5.5in]{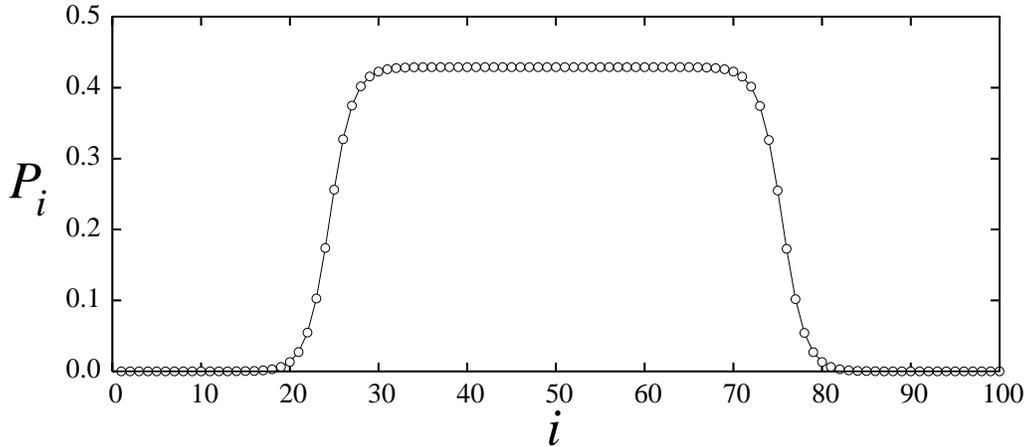}
\caption{Order-parameter profile of the isotropic-nematic interface in the MF
theory for the Lebwohl-Lasher model.}
\label{OPprofIN}\end{figure}

The plate-isotropic, $\gamma_{\hbox{\tiny SI}}$, and plate-nematic, 
$\gamma_{\hbox{\tiny SN}}$, surface tensions have also been calculated 
{in a range of values of the surface coupling constant $\epsilon_s$.
Depending on the value of $\epsilon_s$, different wetting regimes are
obtained. For $\epsilon_s\le 0$ the wall-nematic has an infinitely thick 
isotropic layer adsorbed on the wall, i.e. 
$\gamma_{\hbox{\tiny SN}}=\gamma_{\hbox{\tiny SI}}+\gamma_{\hbox{\tiny IN}}$,
which corresponds to complete wetting of the surface-nematic interface
by the isotropic phase 
(cf. simulations results of Ref. \cite{Doug}). In the case
$\epsilon_s\agt 0.43\epsilon$ the surface tensions satisfy
$\gamma_{\hbox{\tiny SI}}=\gamma_{\hbox{\tiny SN}}+
\gamma_{\hbox{\tiny IN}}$, implying complete wetting by the nematic
phase of the surface-isotropic interface. In the interval
$0<\epsilon_s\alt 0.43\epsilon$ a partial wetting situation arises.
Fig. \ref{Wet} summarises these results.

Finally, we note that the sign of the surface-tension difference 
$\gamma_{\hbox{\tiny SI}}-\gamma_{\hbox{\tiny SN}}$ depends on $\epsilon_s$.
It turns out that both $\gamma_{\hbox{\tiny SN}}$ and
$\gamma_{\hbox{\tiny SI}}$ decrease with $\epsilon_s$, but
$\gamma_{\hbox{\tiny SI}}-\gamma_{\hbox{\tiny SN}}<0$ for
$\epsilon_s<0.219\epsilon$ and $\gamma_{\hbox{\tiny SI}}-\gamma_{\hbox{\tiny SN}}>0$ for
$\epsilon_s>0.219\epsilon$.
When $\epsilon_s=0.219\epsilon$ we have
$\gamma_{\hbox{\tiny SI}}-\gamma_{\hbox{\tiny SN}}=0$; this case lies of course
in the regime of partial wetting (Fig. \ref{Wet}).
}

%For the case $\epsilon_s=0$ we have $\gamma_{\hbox{\tiny SI}}=0$. 
%It turns out that the wall-nematic interface has an infinitely thick 
%isotropic layer adsorbed on the wall, i.e. $\gamma_{\hbox{\tiny SN}}=
%\gamma_{\hbox{\tiny IN}}$, which corresponds to complete wetting by the 
%isotropic phase (cf. simulations results of Ref. \cite{Doug}).
%In the case $\epsilon_s=\epsilon$, which corresponds to the system studied 
%in the present work, the wall is wet by the nematic phase. Fig. \ref{Wet} shows the 
%balance between surface tensions as a function of $\epsilon$. In this case we see that,
%when $\epsilon^*=\epsilon^*_{\hbox{\tiny IN}}$ (with $\epsilon^*\equiv\epsilon/kT$ and
%$\epsilon^*_{\hbox{\tiny IN}}\equiv\epsilon/kT_{\hbox{\tiny IN}}$), 
%Young's law for complete nematic wetting, 
%$\gamma_{\hbox{\tiny SI}}=\gamma_{\hbox{\tiny SN}}+\gamma_{\hbox{\tiny IN}}$, is satisfied;
%{for $\epsilon^*<\epsilon^*_{\hbox{\tiny IN}}$ (or $T>T_{\hbox{\tiny IN}}$)
%there is partial wetting of the substrate by the nematic phase.}
%For different values of $\epsilon_s$ the wetting behaviour may change.

\begin{figure}[h]
\includegraphics[width=4.5in]{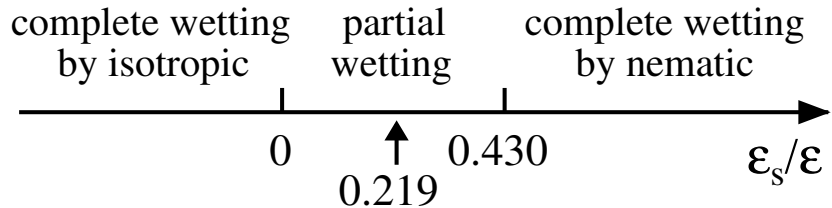}
\caption{Wetting regime of the Lebwohl-Lasher model in mean-field theory.
For $\epsilon_s\le 0$ the substrate is wet by the isotropic phase. For
$\epsilon_s\agt 0.430\epsilon$ the substrate is wet by the nematic phase.
In between a partial-wetting regime occurs. The arrow indicates the case 
where the surface-nematic and surface-isotropic interfaces are equal, which
occurs at $\epsilon_s=0.219\epsilon$.}
%Surface tensions for the case $\epsilon_s^*=1$ as a function
%\caption{Surface tensions for the case $\epsilon_s^*=1$ as a function
%of coupling strength relative to value for bulk transition, as obtained from MF
%theory. Surface tensions are
%given in scaled units $\gamma^*=\gamma\epsilon/a^2$.}
\label{Wet}\end{figure}

\section{MACROSCOPIC ANALYSIS}
\label{apC}

In order to understand the behaviour of the transition line in the confined system, 
one can use a macroscopic analysis involving capillary and elastic forces and derive
a modified Kelvin equation. This is valid whenever the transition is of first order.
The shift in transition temperature $T(h)$ with respect to the bulk temperature
$T_{\hbox{\tiny IN}}$ can be obtained by writing the free energies 
of the confined isotropic and nematic phases. For a relative temperature 
$\Delta T(h)=T(h)-T_{\hbox{\tiny IN}}$ but small 
in absolute value compared to $T_{\hbox{\tiny IN}}$, we have, for the confined
isotropic phase {\cite{Sluckin}}:
\begin{eqnarray}
\frac{F_{\hbox{\tiny I}}}{A}=\gamma_{\hbox{\tiny SI}}^{(1)}+
\gamma_{\hbox{\tiny SI}}^{(2)}+
\left(f_{\hbox{\tiny I}}^{\hbox{\tiny (bulk)}}-s_{\hbox{\tiny I}}\Delta T\right)h+
f_s,
\end{eqnarray}
where $f_{\hbox{\tiny I}}^{\hbox{\tiny (bulk)}}$ is the bulk free-energy density
of the isotropic phase, $s_{\hbox{\tiny I}}$ the entropy density of the isotropic
phase at IN coexistence, {and $f_s(h)$ the free energy per unit area of the 
step interface. $f_s(h)$ is only appreciable for small $h$, i.e. when the two
nematic films are in close contact, and we neglect it here.}

Now the nematic phase is assumed to consist of a linearly-varying director tilt,
with $\phi=0$ at one wall and $\phi=\pi/2$ at the other. Then an elastic 
contribution $F_{\hbox{\tiny elas}}$ has to be added:
\begin{eqnarray}
\frac{F_{\hbox{\tiny N}}}{A}
=2\gamma_{\hbox{\tiny SN}}+\left(f_{\hbox{\tiny N}}^{\hbox{\tiny (bulk)}}-
s_{\hbox{\tiny N}}\Delta T\right)h+\frac{F_{\hbox{\tiny elas}}}{A}.
\end{eqnarray}
At the transition $F_{\hbox{\tiny N}}=F_{\hbox{\tiny I}}$. Using
$f_{\hbox{\tiny N}}^{\hbox{\tiny (bulk)}}=f_{\hbox{\tiny I}}^{\hbox{\tiny (bulk)}}$,
and solving for $\Delta T$:
\begin{eqnarray}
\Delta T=\frac{2\left(
\gamma_{\hbox{\tiny SN}}-\gamma_{\hbox{\tiny SI}}\right)}
{h\Delta s}+\frac{K\pi^2}{8h^2\Delta s},
\end{eqnarray}
where the elastic energy was written as $F_{\hbox{\tiny elas}}=AhKq^2/2=AhK\pi^2/8h^2$,
{with $q=\pi/2h$},
and $\Delta s=s_{\hbox{\tiny N}}-s_{\hbox{\tiny I}}=s_{\hbox{\tiny N}}<0$ since
$s_{\hbox{\tiny I}}=0$.
At the transition $s_{\hbox{\tiny N}}=-0.418 k a^{-3}$.
Using reduced units $h^*=h/a$, $\gamma^*=\gamma a^{2}/\epsilon$,
$T^*=kT/\epsilon$, $s^*=s a^3/k$, $K^*=Ka/\epsilon=3P^2$ (where $P=0.429$ is the
nematic order parameter at the transition, see Table \ref{Table}), and
$\gamma_{\hbox{\tiny SN}}-\gamma_{\hbox{\tiny SI}}=-\gamma_{\hbox{\tiny IN}}$
(since we are in a wetting situation),
\begin{eqnarray}
\Delta T^*=-\left(\frac{2\gamma_{\hbox{\tiny IN}}^*}
{s_{\hbox{\tiny N}}^*}\right)\frac{1}{h^*}+
\left(\frac{3\pi^2P^2}{8s_{\hbox{\tiny N}}^*}\right)\frac{1}{h^{*2}}=
0.084h^{*-1}-1.630h^{*-2}.
\end{eqnarray}
The first term comes from capillary forces and promotes capillary nematization, whereas 
the second is due to the elastic effects and promotes capillary isotropization.

\section{ELASTIC CONSTANT}
\label{elastic}

Since the interaction energy does not couple the relative position of the spins with
their orientation, there exists no distinction between the three Frank elastic constants \cite{deGennes}
in the Lebwohl-Lasher model, and $K_1=K_2=K_3\equiv K$. Priest \cite{Priest} calculated $K$ using
a molecular-field theory. Here we rederive the result of Priest in the language of density-functional
theory, and show numerical results for the bulk elastic constant $K$. Cleaver and Allen \cite{Doug1}
have obtained the elastic constant by simulation.

We consider a smoothy varying director field corresponding to a distorted director. At each spin site
the director unit vector will point along a different direction, and the excess free energy of the bulk,
distorted nematic, will be:
\begin{eqnarray}
\frac{F_{\hbox{\tiny ex}}[\{f\}]}{kT}&=&
-\epsilon^*\sum_{i=1}^{N}\sum_{j\hspace{0.1cm}\hbox{\tiny{(NN)}}}
\int d\hat{\bm\omega}\int d\hat{\bm\omega}^{\prime}f(\hat{\bm\omega}\cdot\hat{\bm n}_i)
f(\hat{\bm\omega}^{\prime}\cdot\hat{\bm n}_j)
P_2(\hat{\bm\omega}\cdot\hat{\bm\omega}^{\prime}).
\label{A100}
\end{eqnarray}
Here we make explicit the dependence of the distribution function $f$ on the director.
Note that the director need not be the same on each site. 
Now consider a smoothy varying director field corresponding to a distorted director.
We assume that the director rotates about the $y$ axis by an angle $\phi$ (see Fig.
\ref{model}), with a 
value proportional to the distance of the $j$th spin from the $i$th spin
along the $z$ axis. Then, assuming $\phi\ll 1$:
\begin{eqnarray}
\hat{\bm\omega}^{\prime}\cdot\hat{\bm n}_j=\hat{\bm\omega}^{\prime}\cdot{\cal R}_y(\phi)\hat{\bm n}_i
=(\omega_x^{\prime},\omega_y^{\prime},\omega_z^{\prime})\cdot
(\sin{\phi},0,\cos{\phi})=\omega_z^{\prime}
+\omega_x^{\prime}\phi-\frac{\omega_z^{\prime}}{2}\phi^2+\cdots
\end{eqnarray}
where $\hat{\bm n}_i=(0,0,1)$ and ${\cal R}_y(\phi)$ is a rotation matrix about the $y$ axis
through an angle $\phi$. Therefore,
% and $\phi\propto(j-i)/a$ (along $y$).  $a$ is the lattice spacing. 
\begin{eqnarray}
f(\hat{\bm\omega}^{\prime}\cdot\hat{\bm n}_j)&=&f\left(\omega_z^{\prime}
+\omega_x^{\prime}\phi-\frac{\omega_z^{\prime}}{2}\phi^2+\cdots\right)\nonumber\\\nonumber\\
&=&f(\omega^{\prime}_z)+\left[\omega^{\prime}_xf^{\prime}(\omega^{\prime}_z)
\right]\phi +\frac{1}{2}\left[\omega^{\prime 2}_y
f^{\prime\prime}(\omega^{\prime}_z)-\omega_z^{\prime}
f^{\prime}(\omega^{\prime}_z)\right]\phi^2+\cdots
\label{A101}
\end{eqnarray}

\begin{table}
\begin{ruledtabular}
\begin{tabular}{ccccccc}
$T^*$ ($\epsilon^*$) & ${P}$ & ${P}$ 
& $K^*$ & $K^*$ & $K^*/{P}^2$ & $K^*/{P}^2$ \\
 & (sim.) & (theo.) & (sim.) & (theo.) & (sim.) & (theo.)\\
\hline\hline
$0.400$ ($2.500$) & $0.8922$ & $0.9260$ & $2.5290$ & $2.5726$ & $3.177$ & $3.000$\\
$0.750$ ($1.333$) & $0.7672$ & $0.8406$ & $1.9747$ & $2.1200$ & $3.355$ & $3.000$\\
$0.900$ ($1.111$) & $0.7668$ & $0.7901$ & $1.6448$ & $1.8730$ & $3.492$ & $3.000$\\
$1.000$ ($1.000$) & $0.6863$ & $0.7471$ & $1.3103$ & $1.6745$ & $3.594$ & $3.000$\\
$1.080$ ($0.926$) & $0.6038$ & $0.7041$ & $0.8587$ & $1.4872$ & $3.693$ & $3.000$\\
$1.321$ ($0.757$) & $-$ & $0.4290$ & $-$ & $0.5520$ & $-$ & $3.000$\\
\end{tabular}
\end{ruledtabular}
\caption{For different values of scaled temperature $T^*$ or inverse scaled temperature
$\epsilon^*=\epsilon/kT$, values of uniaxial nematic order parameter $P$, scaled
elastic constant $K^*$, and ratio $K^*/P^2$ from both MF theory and MC
simulation \cite{Doug1}. The highest temperature corresponds to the bulk phase transition
in the MF theory.}
\label{Table}
\end{table}

\noindent
Introducing (\ref{A101}) into (\ref{A100}), subtracting the contribution from the undistorted
nematic [which is obtained with $\hat{\bm n}_i=\hat{\bm n}_j=(0,0,1)$], noting that the
linear term in $\phi$ vanishes by symmetry and that the 
ideal free-energy term does not contribute to the difference in free energy between distorted
and undistorted fluid, and going to the continuum by using $a^{-3}\int d{\bm r}\to N$ 
and $\phi\to a\partial_z\phi$, 
one arrives at the following expression for the elastic free energy:
\begin{eqnarray}
\frac{F_{\hbox{\tiny elas}}[\{f\}]}{kT}&=&
-\frac{\bar{z}\epsilon^*}{4a}\!\!\int_V\!\! d{\bm r}\!\!
\int\!\! d\hat{\bm\omega}\!\!\int\!\! d\hat{\bm\omega}^{\prime}
f(\omega_z) \left[\omega^{\prime 2}_y
f^{\prime\prime}(\omega^{\prime}_z)-\omega_z^{\prime}
f^{\prime}(\omega^{\prime}_z)\right]
P_2(\hat{\bm\omega}\cdot\hat{\bm\omega}^{\prime})
\left[\partial_z\phi({\bm r})\right]^2.
\label{exx}
\end{eqnarray}

\begin{figure}[h]
\includegraphics[width=3.5in]{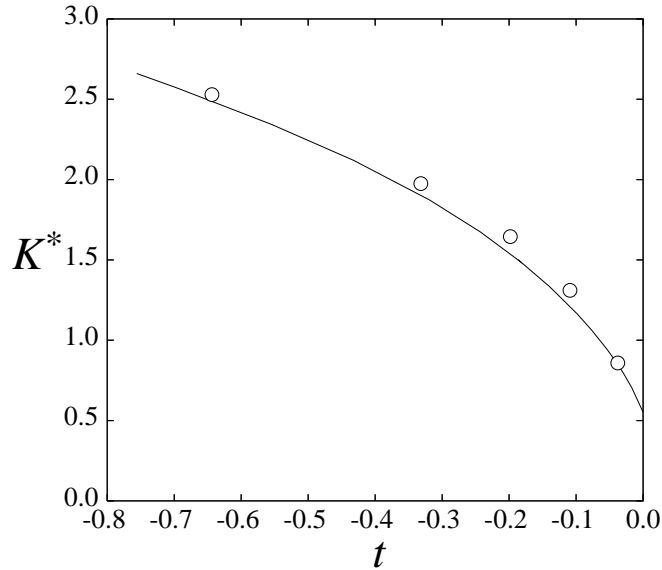}
\caption{Scaled elastic constant $K^*=Ka/\epsilon^*$ for Lebwohl-Lasher model
as a function of relative temperature $t$ (as defined in caption of Fig. \ref{PDasym}).
Circles: simulation results of Cleaver and Allen \cite{Doug}. Open circles:
present MF results.}
\label{Kelas}\end{figure}

\noindent
$\bar{z}$ is an effective coordination number, which is the number of neighbours
of a given one involved in the deformation; since we are rotating about one 
axis, only $2$ out of the $6$ neighbours in the cubic lattice are involved,
so that $\bar{z}=2$. To identify the elastic constant, we
use the expression for the Frank elastic energy \cite{deGennes}. Since our distortion is a bend
mode, we have, with $\hat{\bm n}=(\sin{\phi},0,\cos{\phi})$ and $\phi=qz$ 
(where $q$ is the wavevector of the distortion):
\begin{eqnarray}
F_{\hbox{\tiny elas}}&=&\frac{1}{2}\int_V d{\bm r}
K\left|\hat{\bm n}\times\left(\nabla\times\hat{\bm n}\right)\right|^2=\frac{1}{2}KVq^2,
\end{eqnarray}
where $V$ is the sample volume. 
Comparing with (\ref{exx}), with $\partial_z\phi=q$, we arrive at the expression
\begin{eqnarray}
K^*=-\int d\hat{\bm\omega}\int d\hat{\bm\omega}^{\prime}
f(\hat{\bm\omega}\cdot\hat{\bm z})
P_2(\hat{\bm\omega}\cdot\hat{\bm\omega}^{\prime})
\left[\left(\hat{\bm\omega}^{\prime}\cdot\hat{\bm y}\right)^2
f^{\prime\prime}(\hat{\bm\omega}^{\prime}\cdot\hat{\bm z})
-\left(\hat{\bm\omega}^{\prime}\cdot\hat{\bm z}\right)
f^{\prime}(\hat{\bm\omega}^{\prime}\cdot\hat{\bm z})\right].
\end{eqnarray}
$K^*=Ka/\epsilon$ is the scaled elastic constant. 
{This expression is equivalent to the more general one derived by Poniewierski and
Stecki \cite{PS} in terms of the direct correlation function.}
To calculate $K^*$, we first expand the distribution function using (\ref{expansion}) and
then use the addition theorem of spherical harmonics, so that
\begin{eqnarray}
\int d\hat{\bm\omega}f(\hat{\bm\omega}\cdot\hat{\bm z})
P_2(\hat{\bm\omega}\cdot\hat{\bm\omega}^{\prime})=
\sqrt{\frac{4\pi}{5}}f_{20}P_2(\hat{\bm\omega}^{\prime}\cdot\hat{\bm z}).
\end{eqnarray}
Therefore the elastic constant is:
\begin{eqnarray}
K^*=\left(-\sqrt{\frac{4\pi}{5}}f_{20}\right)
\int d\hat{\bm\omega}^{\prime}\left[
\left(\hat{\bm\omega}^{\prime}\cdot\hat{\bm y}\right)^2
f^{\prime\prime}(\hat{\bm\omega}^{\prime}\cdot\hat{\bm z})
-\left(\hat{\bm\omega}^{\prime}\cdot\hat{\bm z}\right)
f^{\prime}(\hat{\bm\omega}^{\prime}\cdot\hat{\bm z})\right]
P_2(\hat{\bm\omega}^{\prime}\cdot\hat{\bm z}).
\end{eqnarray}
Using again the Legendre expansion of the distribution function, taking 
derivatives, and using a couple of recurrence relations for the Legendre
polynomials, the integral over $\hat{\bm\omega}^{\prime}$ can be calculated
easily, and we obtain the scaled elastic constant:
\begin{eqnarray}
K^*&=&\left(-\sqrt{\frac{4\pi}{5}}f_{20}\right)
\times\left(-3\sqrt{\frac{4\pi}{5}}f_{20}\right)=3{P}^2
\end{eqnarray}
where $P=\left<P_2(\cos{\theta})\right>$ is the uniaxial nematic order parameter,
$P=\left<P_2(\cos{\theta})\right>=f_{20}\sqrt{4\pi/5}$.
Table \ref{Table} presents a comparison of MF theory with MC simulation. At the highest temperature
the MF theory overestimates the elastic constant by almost 75 \%.
The temperature $T^*=1.08$ is $3.8\%$ below the transition temperature from the
simulation. At the same temperature distance from the MF result
($T^*=1.2712$) the comparison is quite good: $K^*=0.8490$ from theory versus
$0.8587$ from simulation. In fact, when plotted versus the variable
$t=(T-T_{\hbox{\tiny IN}})/T_{\hbox{\tiny IN}}$, the two curves are quite close,
see Fig. \ref{Kelas}.


\begin{references}
\bibitem{LL} P. A. Lebwohl and G. Lasher, Phys. Rev. A {\bf 6}, 426 (1972).
\bibitem{LL1} U. Fabbri and C. Zannoni, Mol. Phys. {\bf 58}, 763 (1986).
\bibitem{Zhang} Z. Zhang, M. J. Zuckermann and O. G. Mouritsen, Phys. Rev. Lett. {\bf 69},
2803 (1992).
\bibitem{Zann1} C. Chiccoli, P. Pasini, A. \v{S}arlah, C. Zannoni and
S. \v{Z}umer, Phys. Rev. E {\bf 67}, 050703R (2003).
\bibitem{Zann2} C. Chiccoli, S. P. Gouripeddi, P. Pasini, R. P. N. Murthy,
V. S. S. Sastry and C. Zannoni, Mol. Cryst. Liq. Cryst. {\bf 500}, 118
(2009).
\bibitem{5} P. Palffy-Muhoray, E. C. Garland and J. R. Kelly, Liq. Crys.
{\bf 16}, 713 (1994). 
\bibitem{6} N. Schopohl and T. J. Sluckin, Phys. Rev. Lett. {\bf 59}, 2582
(1987).
\bibitem{9} H. G. Galabova, N. Kothekar and D. W. Allender, Liq. Crys. {\bf 23},
803 (1997).
\bibitem{10} A. \v{S}arlah and S. \v{Z}umer, Phys. Rev. E {\bf 60}, 1821 (1999). 
\bibitem{Bisi} F. Bisi, E. C. Gartland Jr., R. Rosso and E. Virga, Phys. Rev. E {\bf 68},
021707 (2003).

\bibitem{Zappone} {B. Zappone, Ph. Richetti, R. Barberi, R. Bartolino and H. T.
Nguyen, Phys. Rev. E {\bf 71}, 041703 (2005).}

\bibitem{us} D. de las Heras, L. Mederos and E. Velasco,
 Phys. Rev. E {\bf 79}, 011712 (2009).
\bibitem{Paulo} P. I. C. Teixeira, F. Barmes, C. Anquetil-Deck and 
D. J. Cleaver, Phys. Rev. E 79, 011709 (2009).
\bibitem{Frenkel_book}
D. Frenkel and B. Smit, "Understanding Molecular Simulation, From Algorithms
to Applications" (Academic Press, New York, 2002).
\bibitem{Kunz}
H. Kunz and G. Zumbach, Phys. Rev B {\bf 46}, 662 (1992).
\bibitem{Priezjev}
N.V. Priezjev and R. A. Pelcovits, Phys. Rev E {\bf 63}, 062702 (2001).
\bibitem{Wolff}
U. Wolff, Phys. Rev. Lett. {\bf 62}, 361 (1989).
\bibitem{Swendsen}
R.H. Swendsen and J. Wang, Phys. Rev. Lett. {\bf 58}, 86 (1987).
\bibitem{demiguel08}
E. de Miguel, \emph{J. Chem. Phys.} {\bf 129}, 214112 (2008)

\bibitem{Doug} D. J. Cleaver and M. P. Allen, Mol. Phys. {\bf 80}, 253 (1993).

\bibitem{Todos} M. M. Telo da Gama, P. Tarazona, M. P. Allen and R. Evans,
Mol. Phys. {\bf 71}, 801 (1990).
\bibitem{Todos1} M. M. Telo da Gama and P. Tarazona, Phys. Rev. A {\bf 41}, 1149 (1990).
\bibitem{Todos2} P. G. Ferreira and M. M. Telo da Gama, Physica A {\bf 179}, 179 (1991).

\bibitem{Tim} P. Shukla and T. J. Sluckin, J. Phys. A {\bf 18}, 93 (1985).

\bibitem{Landau_book}
D.~P.~Landau and K.~Binder, {\it A Guide to Monte Carlo Simulations in 
Statistical Physics}, 2nd edition (Cambridge University Press, Cambridge 2005).

\bibitem{Salas_2000}
J. Salas and A.D. Sokal, J. Stat. Phys., {\bf 98}, 551 (2000).

\bibitem{BKT1}
V.~Berezinskii,
  Sov. Phys.-JETP, {\bf 34}, 610 (1972).

\bibitem{BKT2}
J.~M. Kosterlitz, D.~J. Thouless, J. Phys. C: Solid State Phys.  {\bf 5}, L124(1972).

\bibitem{Tomita}
Y.~Tomita and Y.~Okabe, Phys. Rev. B {\bf 65}, 184405 (2002).


\bibitem{AML}
N.G. Almarza, C. Mart\'{\i}n and E. Lomba, Phys. Rev. E {\bf 82}, 011140 (2010). 

\bibitem{deGennes} P. G. de Gennes and J. Prost, {\it The physics of liquid crystals}
(Clarendon Press, Oxford, 1993).

\bibitem{Priest} R. G. Priest, Mol. Cryst. Liq. Cryst. {\bf 17}, 129 (1972).

\bibitem{Doug1} D. J. Cleaver and M. P. Allen, Phys. Rev. A {\bf 43}, 1918 (1991).

\bibitem{PS} {A. Poniewierski and J. Stecki, Mol. Phys. {\bf 38}, 1931 (1979).}

\bibitem{Sluckin} {A. Poniewierski and T. J. Sluckin, Liq. Crys. {\bf 2}, 281 (1987).}

\end{references}
\end{document}